\documentclass[]{pasj02} 
\usepackage[switch,mathlines]{lineno} 
\usepackage{natbib} 
\usepackage{siunitx}
\usepackage{tikz}
\usepackage{lscape}
\usetikzlibrary{positioning,shapes.geometric,shapes.misc}
\usepackage{multirow}
\usepackage{makecell}

\jyear{2025}
\Received{}
\Accepted{}

\begin{document} 

\title{Meta-heuristic design of a light-weight homologous backup structure of the primary reflector for the Large Submillimeter Telescope}

\author{
 Chihiro \textsc{Imamura},\altaffilmark{1}\altemailmark \email{imamura@a.phys.nagoya-u.ac.jp} \orcid{0009-0007-7963-7344}
 Yoichi \textsc{Tamura},\altaffilmark{1}\orcid{0000-0003-4807-8117}
 Hiroaki \textsc{Kawamura},\altaffilmark{2}\orcid{0000-0001-8703-3734}
 Toshiaki \textsc{Kimura},\altaffilmark{2}\orcid{0000-0001-8570-1073}
 Akio \textsc{Taniguchi},\altaffilmark{3}\orcid{0000-0002-9695-6183}
 and
 Mikio \textsc{Kurita}\altaffilmark{4}\orcid{0009-0002-8262-4150}
}
\altaffiltext{1}{Department of Physics, Graduate School of Science, Nagoya University, Furo-cho, Chikusa-ku, Nagoya, Aichi 464-8602, Japan}
\altaffiltext{2}{Graduate School of Design and Architecture, Nagoya City University Kita-Chikusa, Chikusa-ku, Nagoya, Aichi 464-0083, Japan}
\altaffiltext{3}{Kitami Institute of Technology, 165 Koen-cho, Kitami, Hokkaido, 090-8507, Japan}
\altaffiltext{4}{Department of Astronomy, Kyoto University, Kitashirakawa-oiwake-cho, Sakyo-ku, Kyoto, Kyoto 606-8502, Japan}



\maketitle

\begin{abstract}
The development of large-aperture submillimeter telescopes, such as the Large Submillimeter Telescope (LST) and the Atacama Large Aperture Submillimeter Telescope (AtLAST), is essential to overcome the limitations of current observational capabilities in submillimeter astronomy. These telescopes face challenges related to maintaining high surface accuracy of the main reflector while minimizing the weight of the telescope structure. This study introduces a genetic algorithm (GA)-based structural optimization, previously applied in related works, to \SI{50}{\meter}-class backup structures (BUSes) with a variable focal position, addressing the challenge of achieving both lightweight construction and high surface accuracy through the consideration of homologous deformation. We model the BUS as a truss structure and perform multi-objective optimization using a GA. The optimization process considers two structures: axisymmetric and non-axisymmetric between the top and bottom. The optimization aims to find structures that simultaneously minimize the maximum stroke length of actuators and the mass of the BUS under practical constraints. The optimized structures show improved surface accuracy, primarily due to the minimization of the maximum actuator stroke length, and reduced weight, both achieved under the imposed constraints. Notably, we find a homologous BUS solution that achieves a surface error of down to $\sim$ \SI{5}{\micro\meter} RMS with a tiny portion of the truss nodes being actively controlled. The results highlight the potential of GA-based optimization in the design of next-generation submillimeter telescopes, suggesting that further exploration of non-axisymmetric structures could yield even more effective solutions. Our findings support the application of advanced optimization techniques to achieve high-performance and cost-effective telescope designs.
\end{abstract}


\section{Introduction} \label{sec:introduction}

In the pursuit of understanding the universe, submillimeter wave observations have become indispensable. These wavelengths reveal crucial information on the cold, distant universe that is often obscured in other spectra. However, current observational capabilities are constrained by the limited number of \SI{10}{\meter} class submillimeter telescopes like the Atacama Pathfinder Experiment \citep[APEX,][]{Gusten2006}, the Atacama Submillimeter Telescope Experiment \citep[ASTE,][]{Ezawa2004}, the Caltech Submillimeter Observatory \citep[CSO,][]{Phillips2007}, which is currently undergoing relocation and refurbishment for the use as the Leighton Chajnantor Telescope \citep[LCT,][]{Vial2020}, the Greenland Telescope \citep[GLT,][]{Chen2023}, the Heinrich Hertz Telescope \citep[HHT,][]{Baars1999}, and the James Clerk Maxwell Telescope \citep[JCMT,][]{Hills1985} or large-scale submillimeter interferometers like the Atacama Large Millimeter/submillimeter Array \citep[ALMA,][]{Wootten2009}, the Northern Extended Millimeter Array \citep[NOEMA,][]{Guilloteau1992} and the Submillimeter Array \citep[SMA,][]{Ho2004}. While these instruments have made significant contributions to radio astronomy, more is needed to meet the growing demands for higher sensitivity, wider fields of view, and broadband observations.

To address the current absence of submillimeter telescopes that offer a wide field of view, high sensitivity, and high angular resolution, the development of large-aperture single-dish submillimeter telescopes, such as the Large Submillimeter Telescope \citep[LST,][]{Kawabe2016}, the Atacama Large Aperture Submillimeter Telescope \citep[AtLAST,][]{Klaassen2020}, and the \SI{60}{\meter} submillimeter telescope in China \citep[][]{Lou2020}, have been proposed. These next-generation telescopes aim to significantly enhance observational capabilities by providing higher sensitivity and larger fields of view. Ongoing studies to realize these telescopes have already yielded remarkable results, including advancements in millimeter adaptive optics for LST \citep[e.g.,][]{Tamura2020, Nakano2022}, detailed design studies for AtLAST \citep[e.g.,][]{Mroczkowski2025, Gallardo2024, Puddu2024, Reichert2024}, and the optimization of the \SI{60}{\meter} submillimeter telescope structure, including its topology, using genetic algorithms (GA) \citep[][]{Gao2022}. However, more developments are still required to realize these large aperture submillimeter telescopes.

Maintaining the accuracy of the primary reflector surface is a major concern, as a telescope itself and environment degrade the optical performance by deforming the structure. This applies especially to the proposed large aperture submillimeter telescopes because of its large structure, required high precision of the primary reflector, and no astrodome. Historically, various techniques have been developed to maintain high surface accuracy, particularly to mitigate deviation of the surface from the ideal shape due to deformation. 
\citet{vonHoerner1967} introduced the theory of homologous deformation for the backup structure (BUS) of telescopes. This theory allows the deformed surface to remain homologous to the original shape, preserving the telescope's performance. Based on this research, the main reflector support structure of the Effelsberg Radio Telescope~\citep{Wielebinkski1970} was designed using homologous principles, enabling large-aperture telescopes such as the IRAM \SI{30}{\meter} Telescope~\citep{Baars1987}, the Large Millimeter Telescope~\citep{Hughes2020}, and the Nobeyama \SI{45}{\meter} Radio Telescope~\citep{Akabane1982} to perform millimeter observations. Additionally, active surface control systems are also vital for maintaining precise reflector surface, and compensating for deformations caused by environmental factors, gravitational forces, and thermal expansion. This technique is employed by modern large-aperture millimeter telescopes, including the Large Millimeter Telescope~\citep{Hughes2020} and the Sardinia Radio Telescope~\citep{Prandoni2017}. 

The need for lightweight structures is also critical. Lighter telescopes are less costly to construct~\citep{Meinel1982}, reduce carbon footprint~\citep{Wang2024}, and can be maneuvered more quickly~\citep{Kurita2020}. However, achieving a balance between lightness and rigidity is challenging since these two properties contradict each other. 
As a result, submillimeter observations with large-aperture single-dish telescopes remain technically challenging. 

An approach to overcome this limitation is the design of a BUS. A successful BUS design would enable us to simultaneously minimize the deformation and mass of the telescope BUS. A BUS for the submillimeter telescope is usually a truss structure composed of joint and straight elements. The combination of these design variables defines the truss and its performance. The difficulty of designing a large aperture telescope BUS is the huge number of conceivable BUS designs. Larger scale telescope requires more truss elements, resulting in the exponentially increased combinations of potential truss. The optimal BUS design should be chosen from the vast combinations considering constraints such as whether the BUS realizes the various performance requirements of the telescope.

This type of problem is called a combinatorial optimization problem. GA~\citep{Holland1975} is a good solver of the problem and there are a few applications to the telescope design. \cite{Kurita2010} used a multi-objective GA (MOGA) to optimize the mount of the Seimei Telescope~\citep{Kurita2020}, a \SI{3.8}{\meter} segmented mirror optical-infrared telescope. The primary objective was to achieve a structure that is both lightweight and capable of maintaining homologous deformation within limited actuator stroke lengths on the reflector surface. The optimization process successfully produced a rigid, lightweight, and non-axisymmetric mount design, resulting in a highly maneuverable telescope. The optimum telescope structure has a natural frequency of \SI{9.5}{\hertz} by setting a rigid structure in every iteration of the optimization. Currently, most telescopes are structured axisymmetrically except for offset-paraboloidal antennas due to easy design and construction. In general, however, a steerable antenna is asymmetrically deformed by non-axisymmetric load of gravity with respect to the elevation angle (EL) axis. This study underscores the efficacy of using GA for structural optimization in astronomical applications. 

Our study specifically aims to demonstrate the effectiveness of a structural optimization method for the supporting structure of large aperture submillimeter telescopes using genetic algorithms through numerical validation. This design method has the potential to address the degradation of optical performance issues caused by surface error and to realize a non-axisymmetric BUS for a large aperture telescope. As a first step of the study, we optimize a given \footnote{In case the topology is treated as a variable, the solutions obtained by \cite{Shintani2024} could be used as initial values of the optimization problem. Although they present only symmetric structures, their solutions are also available in the optimization of non-axisymmetric BUSes if their design variables are set as shown in section \ref{subsubsec:methods_non-axisymmetric_structure}.} 3D BUS model of a \SI{50}{\meter}-class radio telescope. The major difference from \cite{Kurita2010} lies in the scale of the structure; our target is $\sim10$ times larger. Furthermore, in our case, the optimal BUS design allows for variation in the focal position because larger reflectors tend to deform more significantly. Through this study, we evaluate the efficacy of the method in designing a BUS composed of $\sim10$ times more components than the Seimei Telescope considering homologous deformation. A successful design would offer a significant solution to the challenges faced by large aperture submillimeter telescopes, such as LST and AtLAST. 

In this paper, we present a BUS optimization method with a GA and its demonstrations using a proposed design of LST\footnote{Hereafter we simply refer to a target telescope as the LST although we note that the LST and AtLAST projects are to be merged to form a single project.}. Our scope is the shape and size optimization of the BUS, as the topology is given. In section \ref{sec:methods}, we present our structural optimization method to simultaneously minimize surface accuracy and mass of BUS. In section \ref{sec:results}, we demonstrate the application of the method described in section~\ref{sec:methods} for a structural model of LST BUS. In section \ref{sec:discussion}, the performance of an optimized structure is further evaluated through vibration analysis, and how variations in the objective functions impact the optimization results. In section \ref{sec:conclusion} we summarize the findings of this study.

\section{Methods}\label{sec:methods}

\subsection{Structural Optimization Problem}\label{subsec:structural_optimization_problem}

The primary goal of this study is to find the optimal design of the BUS. It is modeled as a truss structure, with the nodal positions and the cross-sectional areas of the truss elements as design variables. To efficiently search for optimal solutions, the truss elements are grouped, and the same design variables are shared among elements within each group. A multi-objective optimization is employed for the optimization to simultaneously minimize both the maximum stroke length of the actuators on the primary reflector surface and the mass of the BUS. During optimization, constraints are imposed on the maximum focal length and the maximum stress on straight elements. In addition, an evaluation method of an objective function and surface accuracy considering the rigidity of the central hub is introduced. We provide overview of the structural optimization in section~\ref{subsubsec:optimization_problem}. Then, the solver of this problem is described in section~\ref{subsubsec:moga}. Next, the objective functions, the constraints, and the surface accuracy are formulated in detail in section~\ref{subsubsec:formulation}. Finally, another method of evaluating an objective function and surface accuracy is introduced in section~\ref{subsubsec:methods_wo_central_hub}. 

\subsubsection{Optimization Problem}\label{subsubsec:optimization_problem}

\begin{figure}[!b]
    \begin{center}
        \includegraphics[width=8cm]{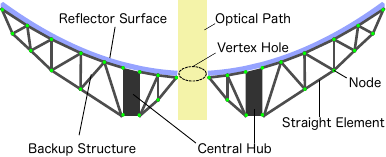}
    \end{center}
        \caption{Schematic diagram of the BUS cross-section. The BUS is composed of nodes and straight elements and is supported by the central hub. {Alt text: Cross-sectionnal view of the BUS in this study.}}
    \label{fig:LST_melco_design_2d}
\end{figure}

\begin{table}
\tbl{BUS components on the optimization.}{%
\begin{tabular}{l p{6cm}}
\hline
Items & Components \\
\hline
Variable &  Joint element, Straight elements \\
Invariable & Not listed above (e.g., Secondary reflector, Supporting structure of secodary reflecotor, Surface panel, Actuator,  Fan) \\
\hline
\end{tabular}}\label{tab:optimized_variables}
\end{table}

\paragraph{Scope of the optimization}
\label{par:scope}
The goal of the structural optimization in this study is to obtain BUSes that are lightweight and capable of achieving an accurate reflector surface. This is achieved by simultaneously minimizing the actuator stroke length on the main reflector surface and the mass of the BUS, under practical constraints. To demonstrate the proposed method, we define the scope of this study as follows. We consider only the gravitational deformation of the BUS as the applied force to simplify the problem. External disturbances, such as gusts, and the loads from the secondary reflector are beyond the scope of this study. Furthermore, the target of the optimization is the BUS of the LST main reflector, the cross-sectional geometry of which is illustrated in figure~\ref{fig:LST_melco_design_2d}. The secondary reflector and its supporting structure are also excluded from the present analysis.

\begin{figure*}[!ht]
    \begin{center}
        \includegraphics[width=18cm]{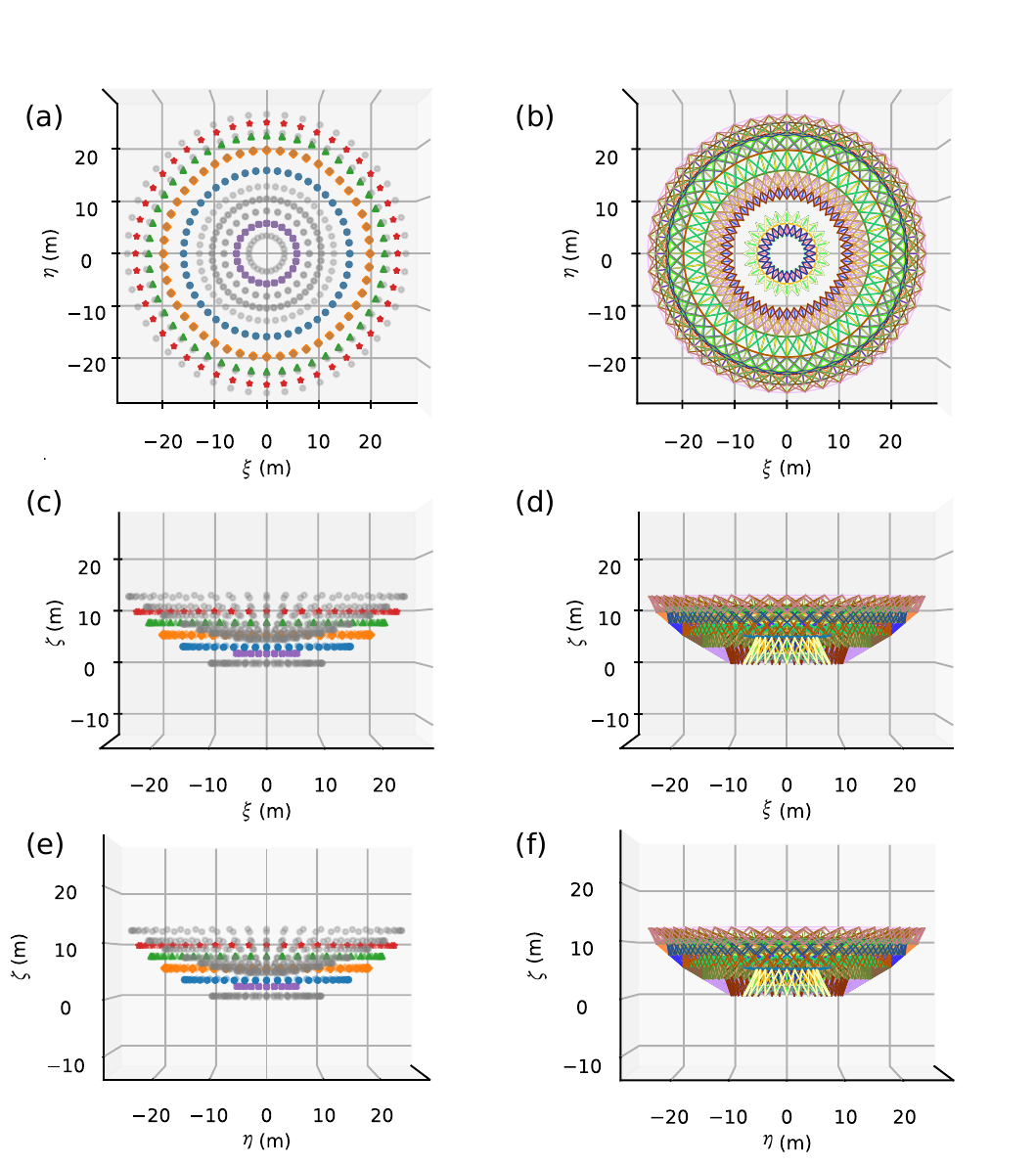}
    \end{center}
    \caption{Initial nodal positions and straight truss elements of the axial symmetric structure in the three view drawings. The origin is set at the bottom of the main reflector and the axes are fixed to the reflector. The reflector rotates on the $\xi$-axis. Colored groups of the truss component in each figure vary its position or cross-sectional area following the corresponding allele. We note that gray points in (a), (c), and (e) are fixed because they are the points on the surface. (a) Top view of nodal positions from $\zeta>0$. (b) Top view of straight truss element positions from $\zeta>0$. (c) Side view of nodal positions from $\eta>0$. (d) Side view of straight truss element positions from $\eta>0$. (e) Side view of nodal positions from $\xi>0$. (f) Side view of straight truss element positions from $\xi>0$. {Alt text: Trihedral figure of axisymmetiric BUS model composed of subfigures labeled from a to f. Nodes and straight elements are circumferentially grouped based on thier radial positions. The Nodes on the central hub or surface are excluded from the design variable.}}
    \label{fig:symmetric_structure}
\end{figure*}

\begin{figure*}[p]
    \begin{center}
        \includegraphics[width=18cm]{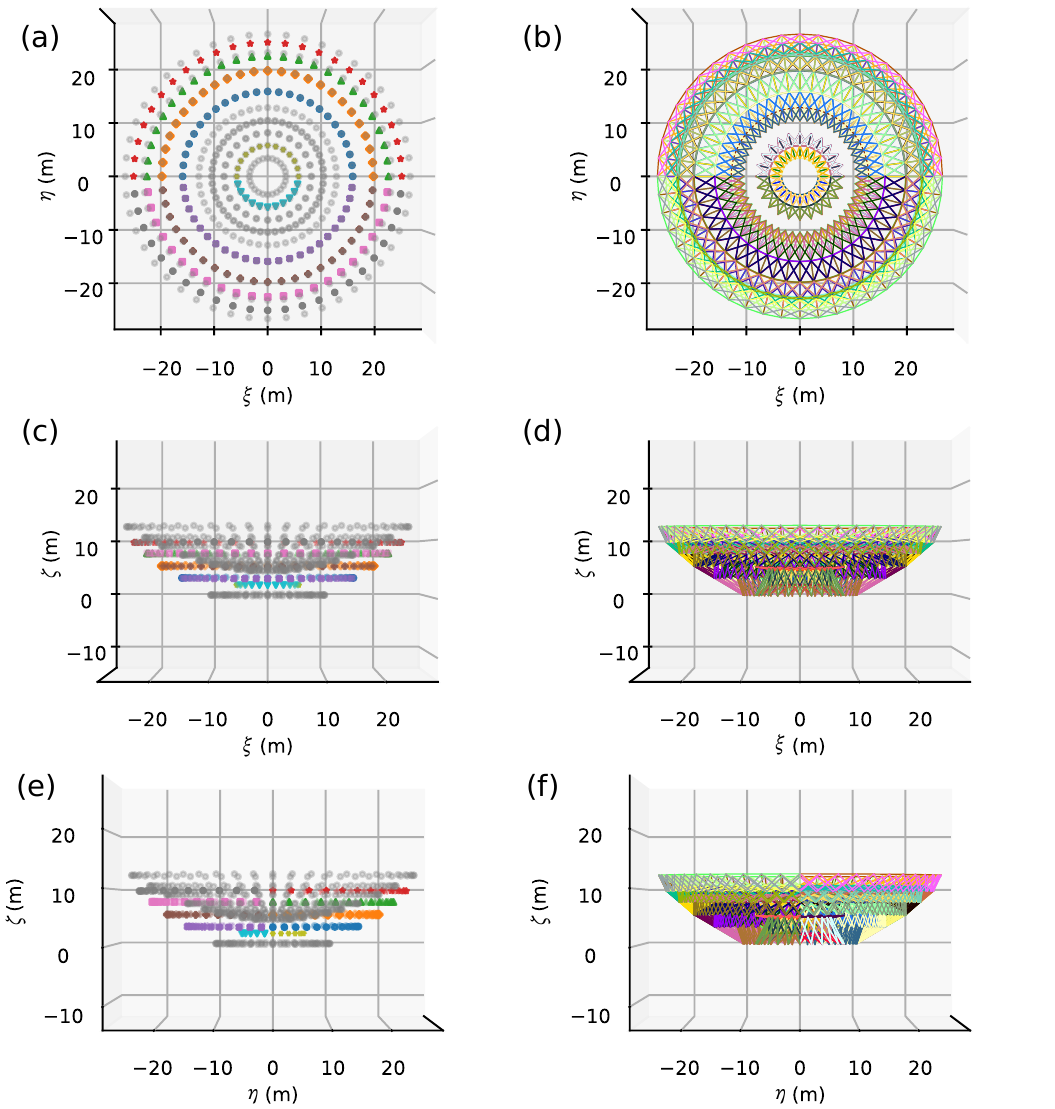}
    \end{center}
    \caption{Initial nodal positions and straight truss elements of the non-axisymmetric structure in the three view drawings. The origin is set at the bottom of the main reflector and the axes are set right-handed and fixed to the reflector. The reflector rotates on the $\xi$-axis. Colored groups of the truss component in each figure vary its position or cross-sectional area following the corresponding allele. We note that gray points in (a), (c), and (e) are fixed because they are the points on the surface. (a) Top view of nodal positions from $\zeta>0$. (b) Top view of straight truss element positions from $\zeta>0$. (c) Side view of nodal positions from $\eta>0$. (d) Side view of straight truss element positions from $\eta>0$. (e) Side view of nodal positions from $\xi>0$. (f) Side view of straight truss element positions from $\xi>0$. {Alt text: Trihedral figure of a non-axisymmetric BUS model composed of subfigures labeled from a to f. Nodes and straight elements are circumferentially grouped based on their radial positions and whether they are located in the upper or lower section of the BUS. The Nodes on the central hub or surface are excluded from the design variable.}}
    \label{fig:asymmetric_structure}
\end{figure*}

\paragraph{Design Variables and their grouping}
\label{par:grouping_of_design_variables}
The BUS is decomposed described in table~\ref{tab:optimized_variables}. Considering the impact on the accurate surface and lightweight BUS, we specifically focus on exploring the optimal design variables of these components, which include 
\begin{itemize}
    \item the nodal displacements of joint elements from their initial positions in the radial ($r$) and vertical ($\zeta$) directions. The nodal displacements are discrete and are chosen from table~\ref{tab:allele}. We do not consider the nodal displacement in the tangential direction for simplicity.
    \item the cross-sectional areas of all the straight elements. The cross-sectional areas of the straight elements are also chosen from table~\ref{tab:allele}.
\end{itemize}
Since the solution space of the design variables is too large ($\sim10^{4800}$ solutions) to explore if we treat them individually, we split the whole straight and joint elements into tens of design variable groups. For example, we make $5$ and $60$ groups for joint element positions and cross-sectional areas of straight elements, respectively, which reduces the solution space by a factor of $\sim10^{4700}$ in the model introduced in section \ref{subsubsec:methods_axisymmetric_model}. 
The design variables are grouped as shown in figures~\ref{fig:symmetric_structure} and \ref{fig:asymmetric_structure} to enhance computational efficiency and simplify the construction. Each colored node group has the same initial displacement in the radial and vertical directions and each colored straight element set has the same cross-sectional area. The relation between these design variables and the allele number, a number that describes a variable in GA, in our optimization process is summarized in table~\ref{tab:allele}. We note that the gray nodes in figures~\ref{fig:symmetric_structure} and \ref{fig:asymmetric_structure} are eliminated from the design variables because they are on the primary reflector surface or the central hub. 

\paragraph{Objective Functions and Constraints}
\label{par:objective_functions_and_constraints}
The optimization aims to minimize both the surface error and the BUS mass under constraints. One objective function is based on the maximum length of the actuators on the primary reflector from an ideal surface. This objective function is set to passively minimize deformation and is calculated taking into account the homologous deformation of the reflector. The BUS is assumed to be adjusted to an ideal paraboloidal surface\footnote{Note that LST/AtLAST adapts the Ritchey-Chr\'{e}tien optics, where a hyperboloid is employed for the primary reflector. See also table~\ref{tab:lst_atlast_specifications}.} at an EL of \ang{50} following the approach of \cite{vonHoerner1975}. The maximum lengths are evaluated at ELs of \ang{85} and \ang{30}, and are converted into a single objective function by averaging their absolute values. We intentionally select the actuator stroke length as one of the objective functions, rather than surface accuracy itself, to actively extend the lifespan of the actuators. Minimizing the actuator stroke length contributes to maintaining the long-term performance of the system while indirectly achieving a high level of surface accuracy discussed in section~\ref{subsec:comparison}. We also evaluate the surface accuracy, the quantity to minimize through the minimization of the maximum actuator stroke length, using obtained actuator stroke length distribution of the optimized structures. The other objective function is the mass of the BUS. This minimization is subject to two constraints: one on the aperture efficiency of the primary reflector, the other on the buckling of the straight elements. When these constraints are violated, penalty terms are multiplied with both objective functions. We note that these penalty term values are eliminated from objective functions unless explicitly stated in section~\ref{sec:results} and section~\ref{sec:discussion}. These objective functions and constraints are formulated in section~\ref{subsubsec:formulation}.

\subsubsection{Multi-objective Genetic Algorithm}\label{subsubsec:moga}

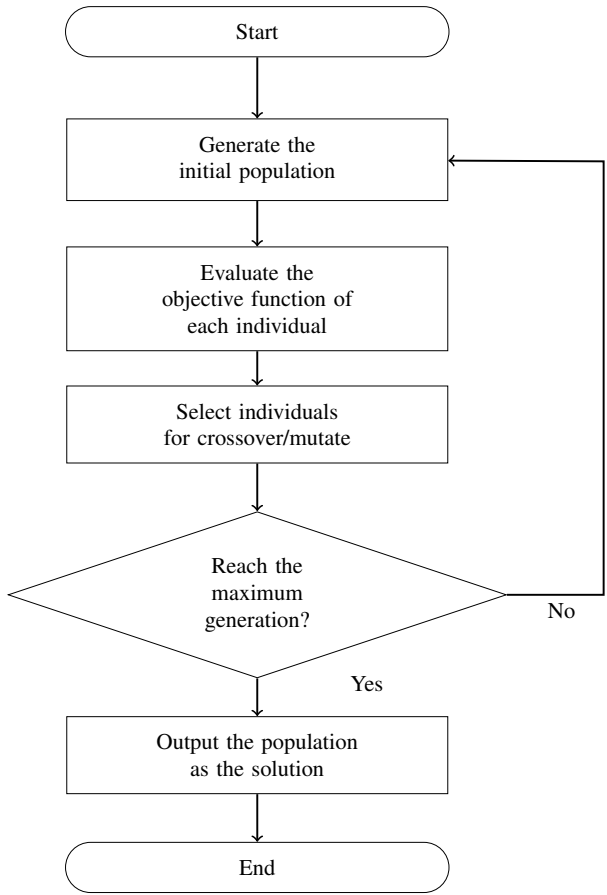
\begin{figure}[t]
    \begin{center}
        \resizebox{8cm}{!}{
    \begin{tikzpicture}
        \tikzset{Terminal/.style={rounded rectangle, draw, text centered, align=center, inner sep=0.25cm, text width=5cm}};
        \tikzset{Process/.style={rectangle, draw, text centered,align=center, inner sep=0.25cm, text width=5cm}};
        \tikzset{Decision/.style={diamond, draw, text centered, aspect=3,align=center, inner sep=0.25cm, text width=2cm}};
        \node[Terminal](a)at (0,0){Start};
        \node[Process, below=1.25 of a.center](b){Generate the \\initial population};
        \node[Process, below=1.25 of b.center](c){Evaluate the \\objective function of \\each individual};
        \node[Process, below=1.25 of c.center](d){Select individuals\\
        for crossover/mutate};
        \node[Decision, below=1.25 of d.center](e){Reach the maximum\\
        generation?};
        \node[Process, below=1.75 of e.center](f){Output the population\\ as the solution};
        \node[Terminal, below=1.25 of f.center](g){End};
        \draw[->, thick] (a) -- (b);
        \draw[->, thick] (b) -- (c);
        \draw[->, thick] (c) -- (d);
        \draw[->, thick] (d) -- (e);
        \draw[->, thick] (e)node[below, xshift=45, yshift=-30]{Yes} -- (f);
        \draw[->, thick] (f) -- (g);
        \draw[->, thick]  (e)node[below, xshift=125]{No}  -- +(5.0, 0.0) -- +(5.0, 6.25) -- (b);
    \end{tikzpicture}
    }
    \end{center}
    \caption{Flow chart of the structural optimization. {Alt text: Flow chart.}}
    \label{fig:flow_chart}
\end{figure}

We adopt the Non-Dominated Sorting Genetic Algorithm ~\citep[NSGA-II,\,][]{Deb2002}, a MOGA solver, for the structural optimization, following the procedure shown in figure~\ref{fig:flow_chart}. This algorithm searches for optimum solutions to multi-objective optimization problems. We assume that a set of design variables defines a BUS just as a chromosome, composed of alleles, defines the traits of a living creature. In MOGA, a set of design variables is called an individual, inspired by the concept of an organism in biology, and each iteration of the algorithm is called a generation, reflecting the stage of biological evolution. 

First, the initial population of individuals is randomly generated. Each individual consists of an array of integers, each component of which is linked to corresponding design variable set. The algorithm then evaluates the objective functions of each individual. Next, the algorithm selects individuals from the population with Non-dominated Sorting~\citep{Deb2002} such that the chosen individuals would not be dominated by any other individuals and be diverse. Crossover involves selected pairs of individuals exchanging portions of their design variables to generate new individuals. This process combines traits from both parents, similar to genetic recombination in biology. Mutation, on the other hand, introduces small random changes to some design variables, ensuring genetic diversity within the population and preventing the algorithm from converging on local minimum solutions. The process repeats, returning to the evaluation step if the maximum number of generations has not been reached; otherwise, the resulting genes represent the optimal solution. We also implement the concept of a ``hall of fame'', wherein we select individuals that outperform all other candidates during the optimization process as valid solutions. In our optimization, the population size is 300, with crossover and mutation probabilities of 0.7 and 0.01, respectively. We perform the optimization in the settings above for 100\,000 iterations, with OpenSeesPy \citep{Zhu2018} for structural calculations, and with DEAP \citep{Fortin2012} for MOGA.

\subsubsection{Formulation of the Objective Functions and Constraints}\label{subsubsec:formulation}
In this section, we detail the formulation of the objective functions and constraints of our optimization to establish clear metrics and boundaries that guide the optimization towards achieving the desired structural performance.

\paragraph{Objective Functions}
The objective function based on the maximum actuator stroke length $D$ is calculated by averaging the absolute values of the maximum deviation from an ideal surface in the unit of \si{\micro\meter} in various ELs of $\theta$. This formulation is based on the assumption that a more accurate surface is expected to require less correction by actuators. We assume that the surface is adjusted to the ideal paraboloidal shape at EL of \ang{50}. We choose two ELs of \ang{85} and \ang{30} to evaluate the objective function as
\begin{align}
    &D_{\theta} = \delta_{\theta, i} - \delta_{50, i}\label{eq:D_el} \\
    &D = \frac{1}{2}\max_i \left(|D_{85}| + |D_{30}| \right),
    \label{eq:objective function_defo}
\end{align}
where $\boldsymbol{\delta}_{\theta}$ is a vector of the actuator stroke lengths of the nodal points on the deformed surface from those of the ideal surface at $\theta$, and $i$ is an index of a nodal point on the surface.
$\boldsymbol{\delta}_{\theta}$ is given by $\boldsymbol{\delta}_{\theta} = \boldsymbol{\sigma}_{\theta} - \boldsymbol{d}_{\theta}$, where $\boldsymbol{\sigma}_{\theta}$ and $\boldsymbol{d}_{\theta}$ are vectors of nodal positions on a deformed surface and ideal surface at the EL, respectively. Accordingly, $D_\theta$ is actuator stroke length at the EL of $\theta$, the maps of which are drawn in figures~\ref{fig:sym_error_dist}, \ref{fig:results_sym}, \ref{fig:results_asym}, \ref{fig:results_wo_hub}, and \ref{fig:results_wo_hub_asym}. We utilize curve fitting to find the ideal surface, taking the coordinates of the focal point and the vertex as free parameters.

The mass-based objective function is the total mass of the straight elements and joint elements of the BUS. We note that all the nodes are counted for calculation of the mass even in the process of optimization ignoring nodes on the central hub.
\begin{equation}
    W = \sum_j^m \rho_j l_j A_j+\sum_k^n w_k,
    \label{eq:mass}
\end{equation}
where $\rho_j$, $l_j$, $A_j$, and $w_k$ are the mass density, length, cross-sectional area of the $j$-th straight element, and the mass of the $k$-th joint element, respectively.

\paragraph{Constraints}
We introduce the constraints on aperture efficiency and stress on straight truss elements. These constraints aim to ensure that the optimized BUS satisfies an optical performance requirement of the telescope and structural safety.

\begin{figure}[tbp]
    \begin{center}
        \includegraphics[width=8cm]{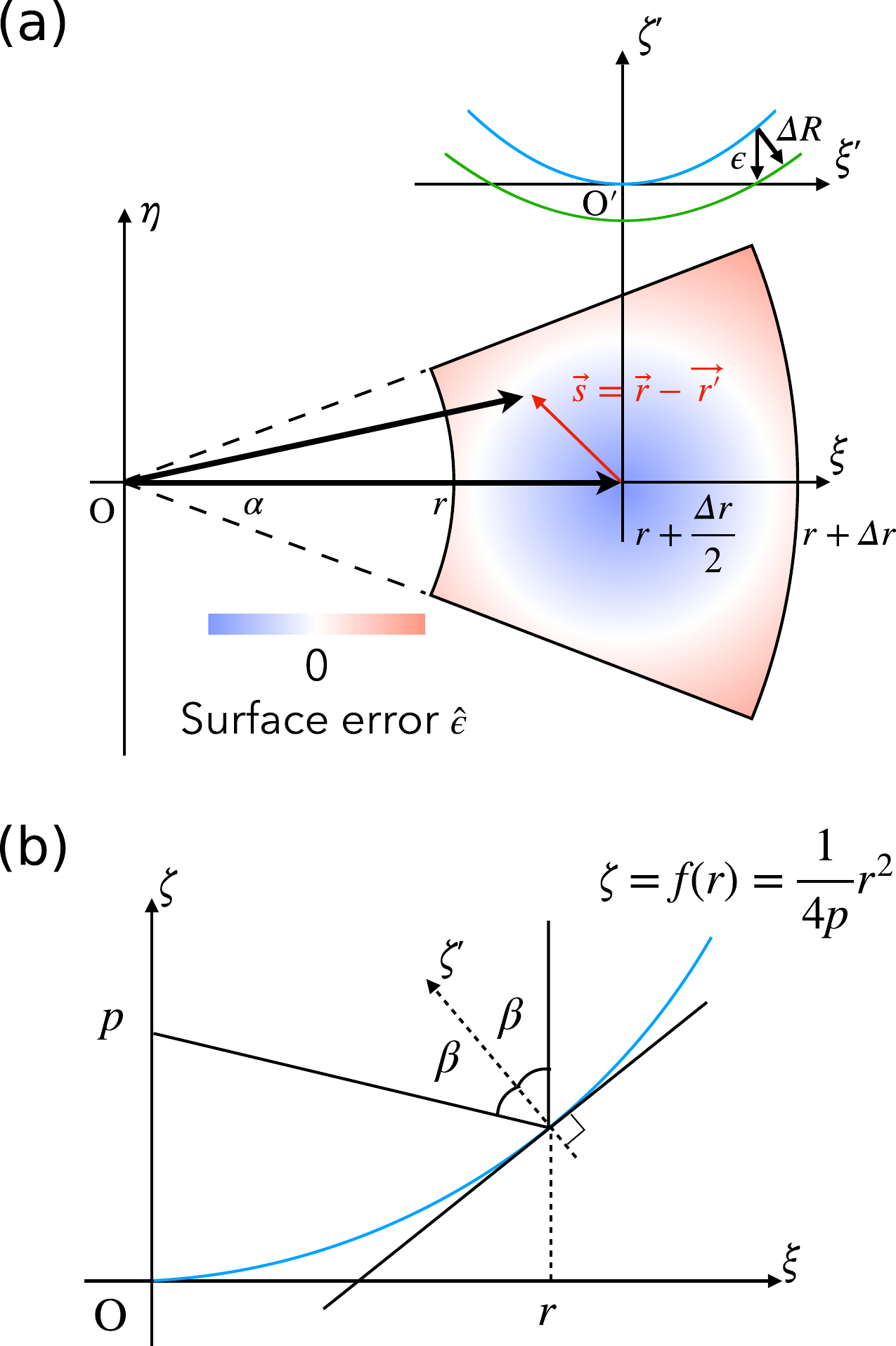}
    \end{center}
    \caption{(a) Schematic diagram of distribution of surface error due to discrepancy between ideal surface and gravitationally deformed surface. $\xi^\prime \eta^\prime \zeta^\prime$-coordinate system in this figure has an origin on the center of the panel, with the $\xi^\prime \eta^\prime$-plane  parallel to the tangent plane of the segmented panel. (b) Optical path and reflection angle in equation~(\ref{eq:delta-epsilon-hat_to_ell}). {Alt text: Two figures showing the geometry in the introduction of power loss owing to deformation of the BUS. The figure consists of subfigures a and b.} }
    \label{fig:constraints_illustration}
\end{figure}

The first one is that the aperture efficiency must be maintained at 90\% or higher after gravitational deformation. We evaluate this constraint as a function of focal length offset and the constraint value is \SI{70}{\milli\meter}. We estimate this value assuming that each segmented mirror is rigid whereas the whole BUS is elastic. This assumption causes a mismatch in focal lengths between the small paraboloidal surfaces created by each segmented mirror and the large paraboloidal surface created by the BUS. First, we convert the change of the focal length offset $\varDelta p$ of a paraboloid $\zeta=\frac{1}{4p}r^2$ to the corresponding change of curvature radius $\varDelta R$ at radius $r$ shown in equation (\ref{eq:delta-p_to_delta-r}).
\begin{equation}
\begin{split}
    \varDelta R(r, p)&=\frac{\partial R(r, p)}{\partial p} \varDelta p \\
    &= \frac{\partial}{\partial p}\left(\frac{\left(1+ f^\prime(r)^2\right)^{3/2}}{\left| f^\prime(r)^\prime \right|}\right)  \varDelta p \\
    &= 2\left(1-\frac{r^2}{2 p^2}\right)\left(1+\frac{r^2}{4 p^2}\right)^{1 / 2} \varDelta p
\end{split}
    \label{eq:delta-p_to_delta-r}
\end{equation}
This discrepancy of curvature causes surface error around $r = r_n + \varDelta r$, where $r_n$ is the internal radius of the segmented panel in the $n$-th tier in the radial direction, and $\varDelta r$ is the length of the segmented panel. Next, we convert $\varDelta R$ into the surface error $\hat{\epsilon}$ as shown in  figure~\ref{fig:constraints_illustration} (a). We introduce the $\xi^\prime \eta^\prime \zeta^\prime$-axes so that $\xi^\prime \eta^\prime$-plane is parallel to the tangent plane of the segmented panel. When the change in curvature is $\varDelta R$, the displacement of the surface in the direction of $\zeta^\prime$ is expressed as 
\begin{equation}
\epsilon(r, \phi)=\frac{\partial \zeta^{\prime}}{\partial R} \Delta R=\frac{R}{\sqrt{R^2-s^2}} \Delta R ,
    \label{eq:delta-r_to_delta-epsilon}
\end{equation}
where $s^2 = \xi^{\prime\,2} + \eta^{\prime\,2}$. 
Considering the correction of the primary reflector so that the mean surface displacement over a surface panel $\langle \epsilon \rangle_{r, \phi} \equiv \frac{\iint\epsilon(r, \phi)r\mathrm{d}r\mathrm{d}\phi}{\iint r\mathrm{d}r\mathrm{d}\phi}$ would be zero, the surface error is expressed as
\begin{equation}
\hat{\epsilon}(r, \phi) = \epsilon(r, \phi) - \langle \epsilon \rangle_{r,\phi} .
    \label{eq:delta-epsilon_to_delta-epsilon-hat}
\end{equation}
The surface error corresponds to a phase error once an observing wavelength is specified. The excess path length is described below using reflection angle $\beta$ in figure~\ref{fig:constraints_illustration} (b).
\begin{equation}
\ell(r, \phi) = \frac{2}{\cos \beta}\hat{\epsilon} = \frac{\sqrt{4p^2 + r^2}}{p}\hat{\epsilon}
    \label{eq:delta-epsilon-hat_to_ell}
\end{equation}
Hence, the corresponding phase error $\delta$ is represented as
\begin{equation}
\delta(r, \phi) = \frac{2\pi}{\lambda}\hat{\ell} = \frac{2\pi}{\lambda}\frac{\sqrt{4p^2 + r^2}}{p}\hat{\epsilon}.
    \label{eq:ell_to_delta}
\end{equation}
Then, the synthesized power over a surface panel in tier $n$ is calculated as 
\begin{equation}
\begin{split}
    E_n&=\iint_{\mathrm{panel }} w(r, \phi) \exp [i \delta(r, \phi)] \mathrm{d} \mathbf{r}^2 \\
    &=\int_{-\alpha}^\alpha \! \int_{r_n}^{r_n+\varDelta r} \! w(r, \phi) \exp [i \delta(r, \phi)] r \mathrm{~d} r \mathrm{~d} \phi ,
\end{split}
\label{eq:delta_to_power}
\end{equation}
where $w(r, \phi)$ is the illumination pattern of the antenna. By applying equation~(\ref{eq:delta_to_power}) to each surface panel, we can evaluate the synthesized power of each panel. Integrating these synthesized powers over the entire surface leads to the received power of the whole antenna as follows:
\begin{equation}
\langle E\rangle=\frac{\sum_{n=1}^N E_n \cdot 2 \pi r_n \varDelta r}{\sum_{n=1}^N 2 \pi r_n \varDelta r} .
\label{eq:E_avg}
\end{equation}
We can evaluate the loss of the aperture efficiency by comparing the normalized received power with one in $\delta = 0$ case,
\begin{equation}
\frac{P}{P_0} = \langle E\rangle \langle E\rangle^* = \left| E \right|^2.
\label{eq:aperture_eff}
\end{equation}

\begin{figure}[!t]
    \begin{center}
        \includegraphics[width=8cm]{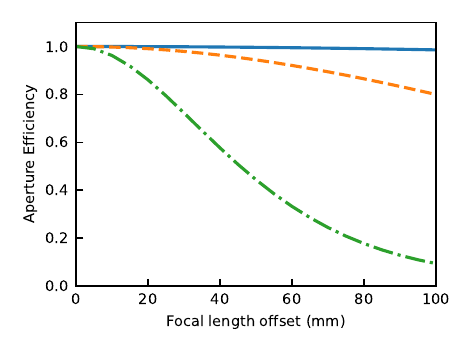}
    \end{center}
    \caption{Aperture efficiency as a function of focal length offset at observing wavelength $\lambda$=\SI{350}{\micro\meter} assuming a uniform illumination pattern ($w$=1) and an aperture of \SI{50}{\meter}. The solid blue line corresponds to a segmented mirror edge length of \SI{1}{\meter}, the dashed orange line to \SI{2}{\meter}, and the dash-dotted green line to \SI{4}{\meter}. {Alt text: Line graph showing the comparison of aperture efficiency as a function of focal length offset in three different segmented mirror sizes.}}
    \label{fig:delta_p_vs_eff}
\end{figure}

We impose the constraint of $\frac{P}{P_0} \ge 90\%$. In order to determine the corresponding limiting focal length offset, we assumed the following factors: the shortest operational wavelength of the LST/AtLAST $\lambda = \SI{350}{\micro\meter}$, the length of a side of surface panels $\varDelta r = \SI{2}{\meter}$ (dashed orange line in figure~\ref{fig:delta_p_vs_eff}), and the uniform illumination pattern $w(r, \phi)=1$. Figure~\ref{fig:delta_p_vs_eff} describes the relation between $P/P_0$ and $\varDelta p$ on the assumption above, which results in the constraint of the focal length offset $\varDelta p \le \SI{70}{\milli\meter}$. Hence, the constraint value is $\varDelta p_\mathrm{lim} = \SI{70}{\milli\meter}$. In addition, we define the ratio of calculated and constraint values as follows:
 \begin{equation}
     \gamma_\mathrm{focal} = \frac{\varDelta p}{\varDelta p_\mathrm{lim}},
 \end{equation}
 where $\varDelta p$ is the actual focal length offset. This value is rewritten as $\gamma_{\mathrm{focal}, \theta}$ considering the dependency on EL.

The second constraint is that the axial stress of each straight truss element does not exceed the limit value specified by the allowable stress design method. We calculate the allowable stress with reference to the Japanese structural design code~\citep[in Japanese]{AIJ2019}, in which the allowable strength is determined based on the strength of the standard strength of the material $F$, the design criteria based on the yield strength and the shape of the element. The straight truss elements are connected by pin joints, which limit the types of stress to axial tension and compression. These stresses can be classified into long-term and short-term stresses. We consider both types of stress. Since this study focuses solely on gravitational deformation, only long-term stresses are considered in the analysis.
The allowable stress level $f$ is defined based on the standard strength of the material, which is $F=\SI{235}{\newton\per\square\milli\meter}$ for a steel pipe. In the case of tensile stress and compressive stress, the limit values $f_\mathrm{t},\ f_\mathrm{c}$ are given by equations (\ref{eq:tole_tensile}) and (\ref{eq:tole_compressive}), respectively. 
\begin{align}
\label{eq:tole_tensile}
f_\mathrm{t} &=
\frac{F}{1.5} \\
\label{eq:tole_compressive}
f_\mathrm{c} &=
\begin{cases}
    \frac{\left\{1-0.4(\Lambda / \Lambda_{\mathrm{crit}})^2\right\}}{\nu} F & \text{if } \Lambda \le \Lambda_{\mathrm{crit}} \\
    \frac{0.277}{(\Lambda / \Lambda_{\mathrm{crit}})^2} F & \text{if } \Lambda > \Lambda_{\mathrm{crit}} \\
\end{cases},
\end{align}
where $\Lambda = l_\mathrm{k}/I$ is slenderness ratio, $l_\mathrm{k}$ is buckling length, $I$ is the radius of gyration, $\Lambda_{\mathrm{crit}} = \sqrt{\pi^2 E_\mathrm{Young} / 0.6 F}$ is the critical slenderness ratio, $\nu = 3/2 + 2/3(\Lambda / \Lambda_{\mathrm{crit}})^2$ is the safety factor, and $E_\mathrm{Young}$ is the Young modulus. Hence, the ratio of calculated stress value to constraint value of the $j$-th straight truss element is expressed as equation (\ref{eq:constaint_stress}). 
\begin{equation}
    \gamma_{\mathrm{stress}, j} =
\begin{cases}
    \frac{f_j}{f_{\mathrm{t}, j}} & \mathrm{if}\ f_j \ge 0 \\
    \left|\frac{f_j}{f_{\mathrm{c}, j}}\right| & \mathrm{if}\ f_j < 0 
    \label{eq:constaint_stress}
\end{cases}
\end{equation}
We calculate the ratio for all elements, and their maximum value is used to judge whether the constraint is satisfied or not as expressed in equation (\ref{eq:constraint_value_stress}).
\begin{equation}
    \gamma_\mathrm{stress} = \max_j \gamma_{\mathrm{stress}, j}
    \label{eq:constraint_value_stress}
\end{equation}
This value is also rewritten as $\gamma_{\mathrm{stress}, \theta}$ considering the dependency on EL.

\paragraph{Formulation}
When all constraints are satisfied, the structural optimization problem is formulated as equation~(\ref{eq:formulation_wo_penalty}).
\begin{equation}
\begin{aligned}
    \text{minimize} & \quad \boldsymbol{f}(\boldsymbol{x}, \boldsymbol{A}) = 
    \begin{cases}
        D(\boldsymbol{x}, \boldsymbol{A}) \\
        W(\boldsymbol{x}, \boldsymbol{A})
    \end{cases} \\
    \text{subject to} & \quad \boldsymbol{g}(\boldsymbol{x}, \boldsymbol{A}) \leq 0
\end{aligned}
\label{eq:formulation_wo_penalty}
\end{equation}
In this formulation, $\boldsymbol{x}$ and $\boldsymbol{A}$ represent the initial nodal positions and cross-sectional areas that are selected as design variables. $\boldsymbol{f}$ and $\boldsymbol{g}$ denote the vectors of objective functions and constraints, respectively.

If any of the constraints are violated, the ratio of the maximum value and the constraint value is multiplied with both of the objective functions. In our demonstration, we set the constraint value used in the optimization as 95\% of the constrained value calculated above. The determination of whether the constraints are satisfied is performed for each EL in the evaluation of the objective functions. The multiplication of penalty terms is also performed at the same time. The formulation of this structural optimization problem is, therefore, expressed as follows:
\begin{equation}
\begin{aligned}
    &\mathrm{minimize}& &\boldsymbol{f}(\boldsymbol{x}, \boldsymbol{A})=
    \begin{cases}
    D^\prime(\boldsymbol{x}, \boldsymbol{A}) \\
    W^\prime(\boldsymbol{x}, \boldsymbol{A})
    \end{cases}\label{eq:formulation_w_penalty} \\
\end{aligned}
\end{equation}
\begin{align}
     D^\prime &= \frac{1}{2}\max_i \left(|\delta_{85, i} - \delta_{50, i}| \prod_{\mathrm{j}}\gamma^\prime_{\mathrm{j}, 85} + |\delta_{30, i} - \delta_{50, i}| \prod_{\mathrm{j}}\gamma^\prime_{\mathrm{j}, 30} \right) \\
     \label{eq:D_prime}
    W^\prime &= \frac{1}{2}\left(\sum_j^m \rho_j l_j A_j+\sum_k^n w_k,\right)\left(\prod_{\mathrm{j}}\gamma^\prime_{\mathrm{j}, 85}+\prod_j\gamma^\prime_{\mathrm{j}, 30}\right),
\end{align}
where $\gamma^\prime_{j, \theta}$ represents the penalty term of the constraint of $j$ at the EL of $\theta$ taking safety factor 95~\% into account. The constraints employed in our framework are aperture efficiency and allowable stress, the penalty terms of which are given as  $\gamma^\prime_{j, \theta}$ in equation~(\ref{eq:penalty_term}). 
\begin{equation}
    \gamma^\prime_{j, \theta} =
\begin{cases}
    \gamma_{j, \theta} + 0.05 &\mathrm{if} \quad \gamma_{j, \theta}\ge0.95 \\
    1 &\mathrm{if} \quad \gamma_{j, \theta}<0.95
    \label{eq:penalty_term}
\end{cases}.
\end{equation}

\paragraph{Surface Accuracy}
Surface accuracy in this study is defined as the deviation of the actual reflector surface from the ideal paraboloidal shape measured in terms of the root mean square (RMS) error. This value is calculated considering the reflector surface correction at EL of \ang{50}. This accuracy at an EL of $\theta$ is given by equation~(\ref{eq:surface_accuracy}). 
\begin{equation}
\label{eq:surface_accuracy}
    \varepsilon_{\theta} = \sqrt{\frac{1}{N_\mathrm{node}}\sum_i^{N_\mathrm{node}} \left(\delta_{EL, i}-\delta_{50, i}\right)^2}, 
\end{equation}
where $N_\mathrm{node}$ denotes the number of nodes on the primary reflector surface. We note that the nodes on the central hub are ignored in the evaluation of $\varepsilon_{\theta}$ using the method described in section \ref{subsubsec:methods_wo_central_hub}. Finally, the surface accuracy indicator to evaluate a model considering several ELs $\varepsilon_\mathrm{eval}$  is formulated as equation~(\ref{eq:surface_accuracy_eval}).
\begin{equation}
\label{eq:surface_accuracy_eval}
    \varepsilon_\mathrm{eval} = \sqrt{\frac{1}{2}\left(\varepsilon_{85}^2 + \varepsilon_{30}^2\right)}
\end{equation}

\begin{landscape}
\begin{table}[ptb]
\caption{Relation between alleles in the GA and the corresponding design variables of the structure. In this table, the design variable group for joint elements are numbered in ascending order of radius and named Set$i$ $(i=0, 1, 2, 3, 4)$.}
\begin{tabular}{cccccccccccccc}
\hline
Allele &
   &
  0 &
  1 &
  2 &
  3 &
  4 &
  5 &
  6 &
  7 &
  8 &
  9 &
  10 &
  11 \\ \hline
\multirow{5}{*}{\begin{tabular}[c]{@{}c@{}}Nodal displacement \\in $r$-direction\\ $\varDelta r$ (\si{\meter})\end{tabular}} &
  Set 0 &
  $-1.150$ &
  $-0.941$ &
  $-0.732$ &
  $-0.523$ &
  $-0.314$ &
  $-0.105$ &
  $+0.105$ &
  $+0.314$ &
  $+0.523$ &
  $+0.732$ &
  $+0.941$ &
  $0$ \\
 &
  Set 1 &
  $-2.575$ &
  $-2.175$ &
  $-1.775$ &
  $-1.375$ &
  $-0.975$ &
  $-0.575$ &
  $-0.175$ &
  $+0.225$ &
  $+0.625$ &
  $+1.025$ &
  $1.425$ &
  $0$ \\
 &
  Set 2 &
  $-1.825$ &
  $-1.541$ &
  $-1.257$ &
  $-0.973$ &
  $-0.689$ &
  $-0.405$ &
  $-0.121$ &
  $+0.164$ &
  $+0.448$ &
  $+0.732$ &
  $+1.016$ &
  $0$ \\
 &
  Set 3 &
  $-1.300$ &
  $-1.077$ &
  $-0.855$ &
  $-0.632$ &
  $-0.409$ &
  $-0.186$ &
  $+0.036$ &
  $+0.259$ &
  $+0.482$ &
  $+0.705$ &
  $+0.927$ &
  $0$ \\
 &
  Set 4 &
  $-1.150$ &
  $-0.941$ &
  $-0.732$ &
  $-0.523$ &
  $-0.314$ &
  $-0.105$ &
  $+0.105$ &
  $+0.314$ &
  $+0.523$ &
  $+0.732$ &
  $+0.941$ &
  $0$ \\ \hline
\multirow{5}{*}{\begin{tabular}[c]{@{}c@{}}Nodal displacement \\in $\zeta$-direction\\ $\varDelta z$ (\si{\meter})\end{tabular}} &
  Set 0 &
  $-1.000$ &
  $-0.818$ &
  $-0.636$ &
  $-0.455$ &
  $-0.273$ &
  $-0.091$ &
  $+0.091$ &
  $+0.273$ &
  $+0.455$ &
  $+0.636$ &
  $+0.818$ &
  $0$ \\
 &
  Set 1 &
  $-1.950$ &
  $-1.595$ &
  $-1.241$ &
  $-0.886$ &
  $-0.532$ &
  $-0.177$ &
  $+0.177$ &
  $+0.532$ &
  $+0.886$ &
  $+1.241$ &
  $+1.595$ &
  $0$ \\
 &
  Set 2 &
  $-1.850$ &
  $-1.514$ &
  $-1.177$ &
  $-0.841$ &
  $-0.505$ &
  $-0.168$ &
  $+0.168$ &
  $+0.505$ &
  $+0.841$ &
  $+1.177$ &
  $+1.514$ &
  $0$ \\
 &
  Set 3 &
  $-1.300$ &
  $-1.064$ &
  $-0.827$ &
  $-0.591$ &
  $-0.355$ &
  $-0.118$ &
  $+0.118$ &
  $+0.355$ &
  $+0.591$ &
  $+0.827$ &
  $+1.064$ &
  $0$ \\
 &
  Set 4 &
  $-1.200$ &
  $-0.982$ &
  $-0.764$ &
  $-0.546$ &
  $-0.327$ &
  $-0.109$ &
  $+0.109$ &
  $+0.327$ &
  $+0.546$ &
  $+0.764$ &
  $+0.982$ &
  $0$ \\ \hline
\begin{tabular}[c]{@{}c@{}}Outer diameter\\ $D_\mathrm{out}$ (\si{\milli\meter})\end{tabular} &
  All straight elements &
  $48.60$ &
  $60.50$ &
  $76.30$ &
  $89.10$ &
  $101.6$ &
  $114.3$ &
  $139.8$ &
  $165.2$ &
  $190.7$ &
  $216.3$ &
  $267.4$ &
  $318.5$ \\
Thickness $d$ (\si{\milli\meter}) &
  All straight elements &
  $2.3$ &
  $2.3$ &
  $2.8$ &
  $3.2$ &
  $3.2$ &
  $3.5$ &
  $3.5$ &
  $4.5$ &
  $5.3$ &
  $4.5$ &
  $6.0$ &
  $6.0$ \\ \hline
\end{tabular}
\label{tab:allele}
\end{table}
\end{landscape}

\subsubsection{Optimization Considering Central Hub Rigidity}\label{subsubsec:methods_wo_central_hub}
\begin{figure}[tbp]
    \begin{center}
        \includegraphics[width=8cm]{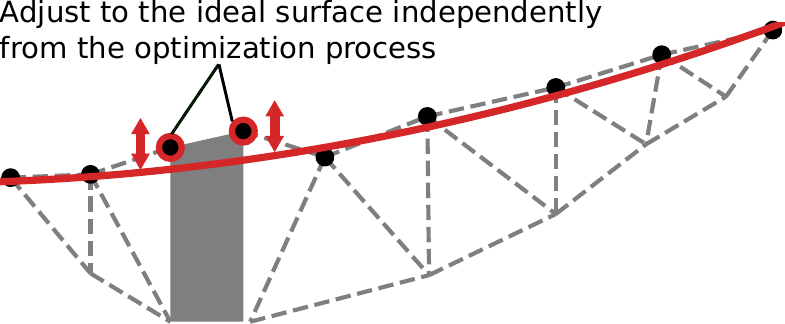}
    \end{center}
    \caption{Conceptual diagram of the evaluation of structural optimization considering the rigidity of the central hub. The rigid hub supports the deformed BUS (dashed gray line), which may hinder homologous deformation. Evaluating the objective function without surface nodes on the hub (nodes: red-outline circles, ideal surface: red curve) can address this issue. The nodes on the central hub can be adjusted to the ideal surface obtained by ignoring them. {Alt text: Cross-sectional view of the BUS, highlighting the necessity of considering the rigidity of the hub.}}
    \label{fig:fit_wo_cntrhub}
\end{figure}

We also attempt the optimization considering the rigidity of the central hub for further minimization of the maximum actuator stroke length and the surface accuracy as well. The BUS we optimize is supported by the rigid central hub, which we regard as an ideal solid body. This rigidity is likely to hinder the homologous deformation of the entire BUS, resulting in low surface accuracy around the hub. Therefore, in this modified method, we treat the nodes on the central hub independent from the other surface nodes. The optimization process ignores the actuator stroke lengths of the nodes on the hub when evaluating the objective function and surface error as shown in figure~\ref{fig:fit_wo_cntrhub}. Then, the nodes on the hub are adjusted to the ideal surface. We note that the deformation around the hub of the optimized structures using this method is expected to be larger than that of the other nodes on the surface. Hence, we also evaluate the actuator stroke length required to correct the deformation around the hub. 

\subsection{Models and Cases Used in the Optimization}\label{subsec:models}
This section introduces the models used in our structural optimization. We provide a basic description of the structure to be optimized in section~\ref{subsubsec:ground_structure}. Sections~\ref{subsubsec:methods_axisymmetric_model} and ~\ref{subsubsec:methods_non-axisymmetric_structure} describe the two models created from this structure. One model is axisymmetric with $\zeta$-axis and the other is differently designed between the upper and lower parts. Each model is optimized using two methods in terms of the hub rigidity described in~\ref{subsubsec:methods_wo_central_hub}. Hence, the structural optimization is performed for these four cases in our study.

\subsubsection{Structure to Be Optimized}\label{subsubsec:ground_structure}

\begin{table}[t]
\caption{Main specifications of the proposed \SI{50}{m}-class single-dish submillimeter telescopes. The 3-D model in this study follows the specifications of LST except for optics. They are excerpted from \cite{Kawabe2016} and \cite{Mroczkowski2025}.}
\label{tab:lst_atlast_specifications}
\begin{center}
    \begin{tabular}{lll}
    \hline
    & LST & AtLAST \\
    \hline
    Optics & \multicolumn{2}{l}{Ritchey-Chr\'{e}tien} \\
    Primary F-number & $0.40$ & $0.35$ \\
    Surface accuracy ($\si{\micro\meter}\mathrm{RMS}$) & $\le 45$ & 20--25 \\
    Diameter of the vertex hole (\si{\milli\meter}) & 6109.115 & 11800 \\ \hline
    \end{tabular}
\end{center}
\end{table}

To demonstrate our structural optimization, we create a 3-D model of the paraboloidal BUS of a \SI{50}{\meter} radio telescope by referring to figures only in \cite{Kawabe2016} (e.g., figures~5, 6, and 7). The original design by \citet{Kawabe2016} aims to realize submillimeter observations up to the observing frequency of \SI{950}{\giga\hertz} with a \SI{50}{\meter} class single-dish telescope. Its main reflector is composed of segmented mirrors, which are supported by a BUS. The main specifications are listed in table~\ref{tab:lst_atlast_specifications}. We follow all the specifications proposed by \citet{Kawabe2016}, except for the optics, where we use a paraboloid instead of a hyperboloid, to balance a simple evaluation of optimized structures and the correct reproduction of the model. 

Figure~\ref{fig:symmetric_structure} is a blueprint of the traced structure. We introduce a 3D Cartesian coordinate system, the origin of which is set at the bottom of the main reflector. The $\xi$-axis is set to be the rotation axis of the EL. This model consists of 3528 straight truss elements and 648 joint elements made of steel. The specification of all the components of the truss is followed by the Japanese Industrial Standards (JIS) G 3444: STK 400, similar to the American Society for Testing and Materials (ASTM) A500: Grade A, Grade B, and Grade C, and the European Norm (EN) 10219: S235. Each straight element has a mass density of \SI{7850}{\kilo\gram\per\cubic\meter}. The joint element is assumed to be small with a negligible size, and its mass is \SI{13.9}{\kilo\gram}. In the radial direction, the structure is considered as a cantilever supported by the rigid, ring-shaped hub with an inner diameter of \SI{15.80}{\meter}, an outer diameter of \SI{20.00}{\meter}, and a height of \SI{5.000}{\meter}. All the straight elements are pin-supported by connecting joint elements, which makes the whole system a truss structure. The structure supports the primary reflector composed of segmented mirrors. We assume the load per area to be \SI{55}{\kilo\gram\per\meter^2} considering the components shown in table~\ref{tab:load}.

\begin{table}[bt]
\caption{Details of the load on a BUS surface.}
\begin{center}
    \begin{tabular}{>{\raggedright\arraybackslash}p{2.5cm} >{\raggedright\arraybackslash}p{2.0cm} >{\raggedright\arraybackslash}p{3.5cm}}
        \hline
        Component        & Mass per area (\si{\kilo\gram\per\meter^2}) & Note \\ \hline
        Surface panel    & 15            & Aluminum\\
        Backboard        & 5.0           & \makecell[l]{CFRP skin,\\Aluminum honeycomb}\\
        Segment          & 20            & Steel\\
        Sun shield panel & 1.0           & Aluminum honeycomb\\
        Actuators        & 3.0           &\\
        Fan              & 1.0           &\\
        Cable            & 10            &\\
        Insulators       & 0.0           & Ignored\\ \hline
    \end{tabular}
\end{center}

\label{tab:load}
\end{table}

\subsubsection{Axisymmetric Model}\label{subsubsec:methods_axisymmetric_model}
The axisymmetric model is set so that every design parameter group would be axisymmetric with $\zeta$-axis. Figure~\ref{fig:symmetric_structure} is the trihedral figure of this model. The structural optimization of the model is carried out with the two methods; one considers the rigidity of the hub described in section~\ref{subsubsec:methods_wo_central_hub} and the other does not. We consider the first case as the fiducial case throughout this paper. This model has $5$ joint element position groups and $60$ cross-sectional area groups of straight elements. Consequently, the total number of sets of design variables for the model is $12^{65}\sim 10^{70}$.

\subsubsection{Non-axisymmetric Model}\label{subsubsec:methods_non-axisymmetric_structure}
In the demonstration of our structural optimization, we also introduce a non-axisymmetric BUS design. The design variable sets are distributed differently for the upper and lower parts of the primary reflector, as shown in figure~\ref{fig:asymmetric_structure}. \citet{Kurita2010} obtained the non-axisymmetric mount of a \SI{3.8}{\meter} class optical-infrared telescope with GA enabling the fast drive of the telescope referred in section~\ref{sec:introduction}. We test the effectiveness of the MOGA-based optimization algorithm on the truss structure with significantly larger ($\sim$3500) truss elements and allow larger homologous deformation. The optimization is performed both with and without considering hub rigidity, as described in section~\ref{subsubsec:methods_axisymmetric_model}. This asymmetry greatly increases the possible combinations to $\sim 10^{140}$.

\section{Results}\label{sec:results}
In this section we present the results of structural optimization and investigate the structural characteristics of the BUS that exhibits the highest surface accuracy in each case. We note that the minimization of surface error shown in this section is done through the minimization of the maximum value of the actuator stroke length as described in section~\ref{par:objective_functions_and_constraints}.

\subsection{Axisymmetric Case (Fiducial Case)}\label{subsec:axially_symmtric}

\begin{figure}[tpb]
    \begin{center}
        \includegraphics[width=8cm]{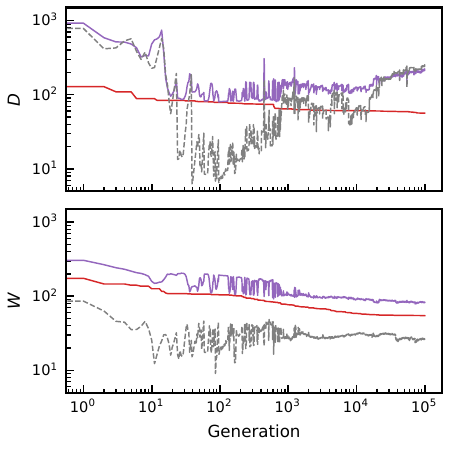}
    \end{center}
    \caption{Statistics in the optimization process of the axisymmetric BUS. The upper panel shows the objective function of the maximum actuator stroke, and the lower panel shows the objective function of the mass of the BUS. The red line represents the minimum value of the solution, the purple line represents the average, and the gray dashed line depicts the standard deviation. {Alt text: Two line graphs.}}
    \label{fig:sym_stats}
\end{figure}

\begin{figure}[tbp]
    \begin{center}
        \includegraphics[width=8cm]{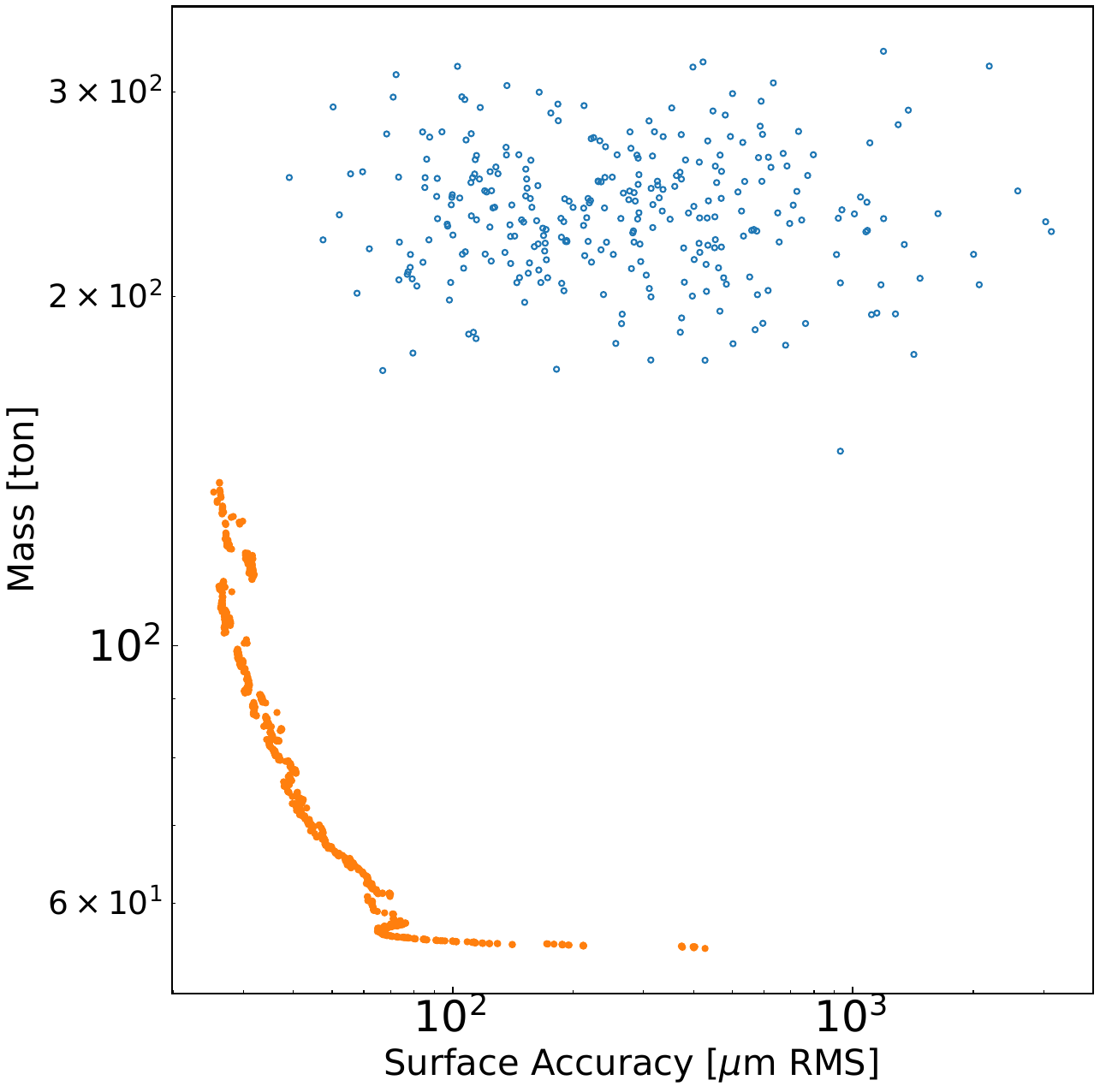}
    \end{center}
    \caption{Randomly generated structures and optimized structures in the surface accuracy and mass of the structure plane in the fiducial case. Blue circles represent randomly generated initial generation and solid orange circles show final generation. {Alt text: Scatter graph comparing surface accuracy and BUS mass between unoptimized and optimized structures.}}
    \label{fig:sym_pareto_acc_mass}
\end{figure}

\begin{figure*}[tbp]
    \begin{center}
        \includegraphics[width=18cm]{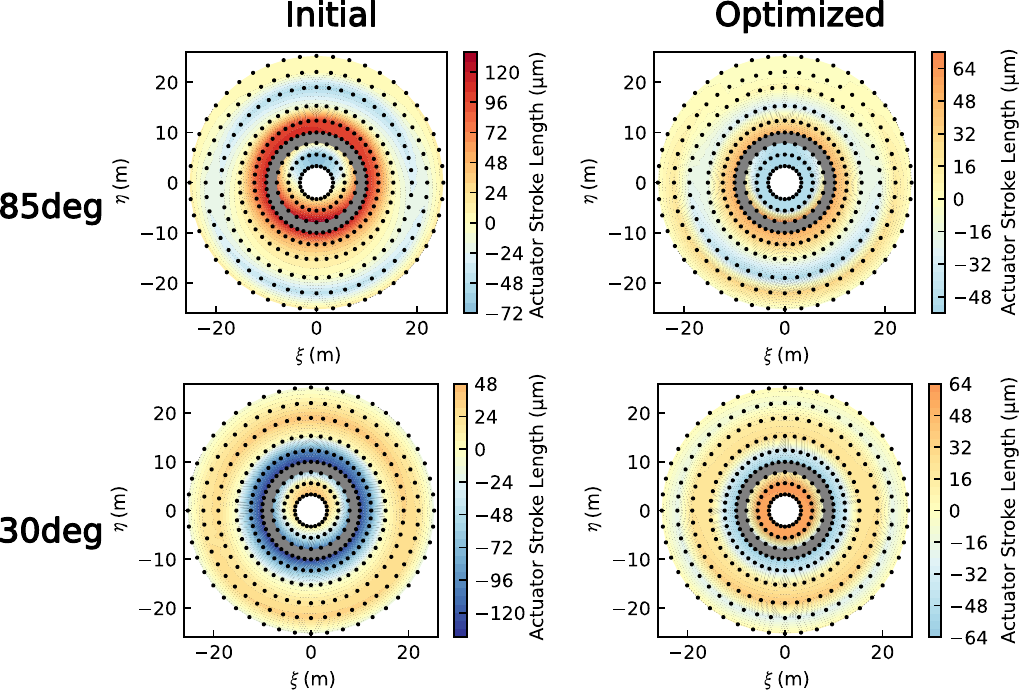}
    \end{center}
    \caption{Distribution of actuator stroke length $D_\theta$ of the most rigid structures in the initial and final generations of the fiducial case at ELs of \ang{85} and \ang{30}, with a common color map scale. Top left: The most rigid structure in the initial generation at EL \ang{85}. Top right: The most rigid structure in the final generation at EL \ang{85}. Bottom left: The most rigid structure in the initial generation at EL \ang{30}. Bottom right: The most rigid structure in the final generation at EL \ang{30}. {Alt text: Maps comparing the deformation of the reflector surface between unoptimized and optimized BUS with four subfigures, illustrating the impact on the optimization.}}
    \label{fig:sym_error_dist}
\end{figure*}

\begin{figure}[tbp]
    \begin{center}
        \includegraphics[width=8cm]{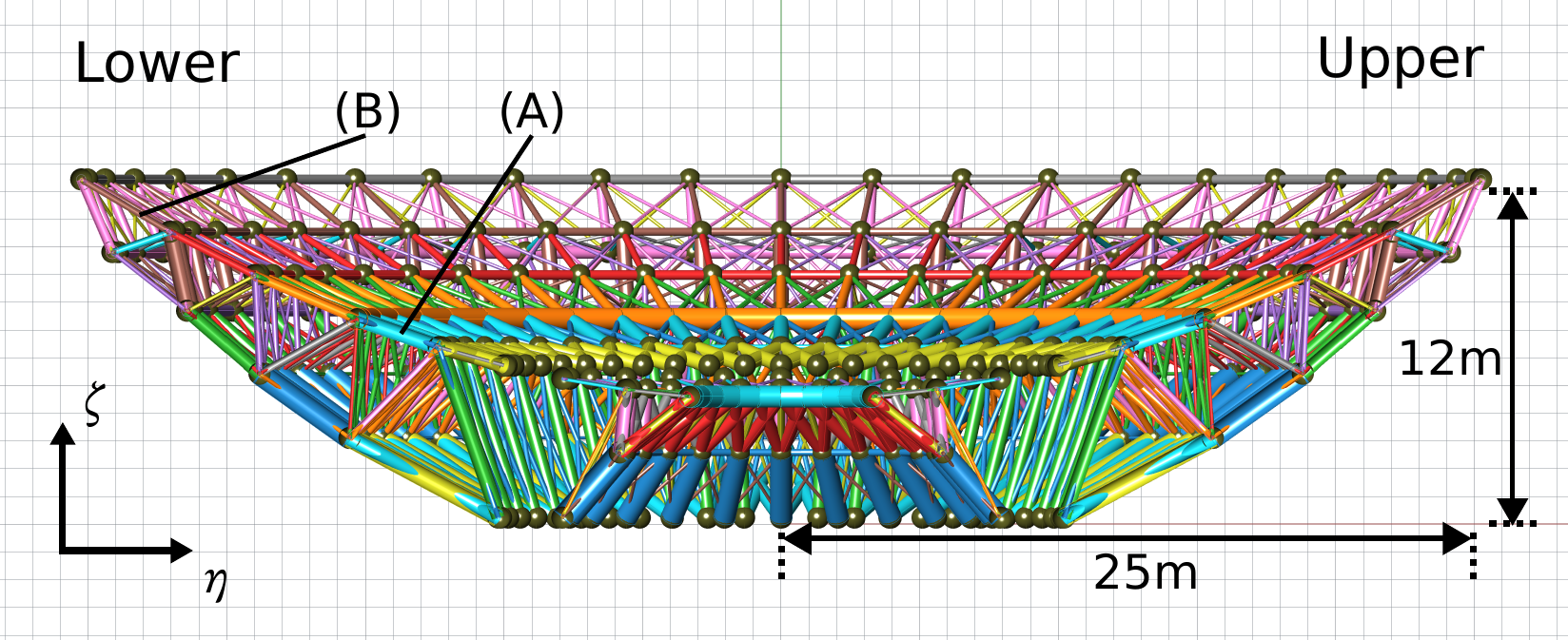}
    \end{center}
    \caption{Cross-sectional view of the optimized structure in the fiducial case at the plane $\xi=0$. The width of the elements in this figure corresponds to the actual cross-sectional area of the optimized structure. The straight elements are colored according to their respective design variable groups. (A) and (B) in the figure highlight examples of elements with large and small cross-sectional areas, respectively. {Alt text: CAD image illustrating the distribution of straight elements in the optimized BUS, showing its cross-section.}}
    \label{fig:optimized_cs}
\end{figure}

\begin{figure*}[tbp]
    \begin{center}
        \includegraphics[width=18cm]{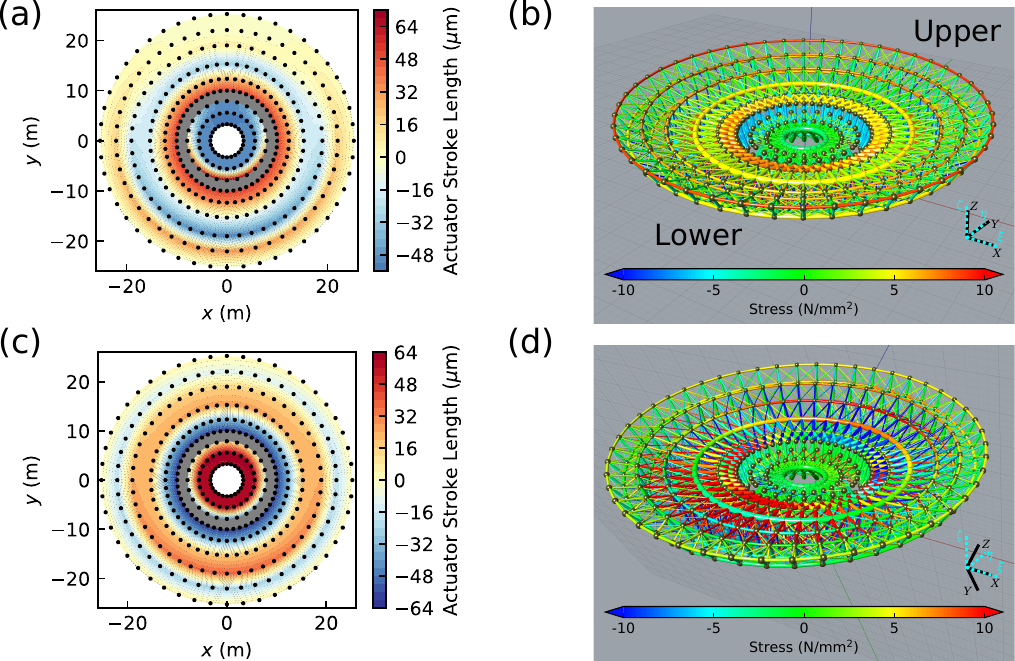}
    \end{center}
    \caption{Maps of actuator stroke length $D_\theta$ and the deformed structures of the optimized structure with the most accurate surface in the axisymmetric (fiducial) case. (a) Distribution of actuator stroke lengths at an EL of \ang{85}. (b) Deformed BUS at an EL of \ang{85}. (c) Distribution of actuator stroke lengths at an EL of \ang{30}. (d) Deformed BUS at an EL of \ang{30}. The dotted lightblue axes of (b) and (d) are the local coordinate fixed to the reflector; The solid black axes are the global coordinate. The thickness of the straight truss elements in panels (b) and (d) is exaggerated, and the deformation of the structure is exaggerated by 1000 times. The color of truss elements in panels (b) and (d) corresponds to the stress value: blue indicates compressive stress, and red indicates tensile stress. The colors of truss elements with compressive or tensile stresses over \SI{10}{\newton\milli\meter^{-2}} are saturated. {Alt text: Maps and CAD images of the BUS that exhibits the highest surface accuracy in the axisymmetric case, with subfigures labeled a to d, illustrating the structural traits of the BUS.}}
    \label{fig:results_sym}
\end{figure*}

First, we validate the convergence of optimization using the statistics of the obtained solutions. Figure~\ref{fig:sym_stats} shows the statistics of the optimization process. Around 100\,000th generation, the minimum values does not change drastically and the standard deviations converge to a finite value. This means that the solutions are sufficiently minimized while their diversity is sustained. 

The surface accuracy and the BUS mass are simultaneously optimized with MOGA as shown in figure~\ref{fig:sym_pareto_acc_mass}. The distribution of physical characteristics in the objective function space is analyzed, particularly looking at maximum actuator stroke length, surface accuracy, and mass. These range from \SI{56.35}{\micro\meter} to \SI{1304}{\micro\meter} for maximum actuator stroke, and from \SI{26.14}{\micro\meter} RMS to \SI{427.4}{\micro\meter} RMS for surface accuracy, from 54.84 tons to 138.0 tons for mass, respectively. Surface accuracy and mass form a distinct trade-off relationship. Additionally, all the optimized structures satisfy both constraints of aperture efficiency and buckling of the truss elements.

We also visually verify the effectiveness of the optimization. Figure~\ref{fig:sym_error_dist} is the comparison between the structures with the most accurate reflector surface in the initial and final generation. An appropriate grouping of design variables enables relatively accurate surfaces even in the initial generation as described in the top left and bottom left panels. The optimization algorithm alleviates the deformation in the entire surface although large error around the central hub remains.

Here we pick up and investigate the BUS that exhibits the highest surface accuracy. Figure~\ref{fig:optimized_cs} shows that the inner elements have larger cross-sectional areas (e.g., element (A) in figure~\ref{fig:optimized_cs}), whereas the outer elements are designed to be lighter (e.g., element (B) in figure~\ref{fig:optimized_cs}). This design contributes to improvements in load support and the overall lightweight nature.
At an EL of \ang{85}, this structure utilizes circumferential elements like hoops, supporting significant tensile or compressive forces as shown in figure~\ref{fig:results_sym} (b). The distribution of actuator stroke length is depicted in the panel (a) of the figure. There is a mode of deformation that is likely to cause defocusing and a large error near the central hub owing to the rigidity around the hub. At an EL of \ang{30}, figure~\ref{fig:results_sym} (d) indicates large tensile and compressive stress in the lower and upper parts, respectively. Figure~\ref{fig:results_sym} (c) shows a similar shape of the actuator stroke length distribution to the panel (a), but the difference in EL changed the sign of the length. 

\subsection{Non-axisymmetric Case}\label{subsec:non-axially_asymmetric}

\begin{figure}[tbp]
    \begin{center}
        \includegraphics[width=8cm]{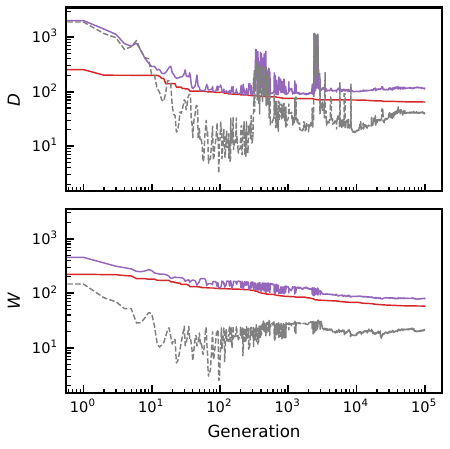}
    \end{center}
            \caption{Statistics in the optimization process of the non-axisymmetric BUS. The upper panel shows the objective function of the maximum actuator stroke, and the lower panel shows the objective function of the mass of the BUS. The red line represents the minimum value of the solution, the purple line represents the average, and the gray dashed line depicts the standard deviation. {Alt text: Two line graphs.}}
    \label{fig:asym_stats}
\end{figure}

\begin{figure}[tbp]
    \begin{center}
        \includegraphics[width=8cm]{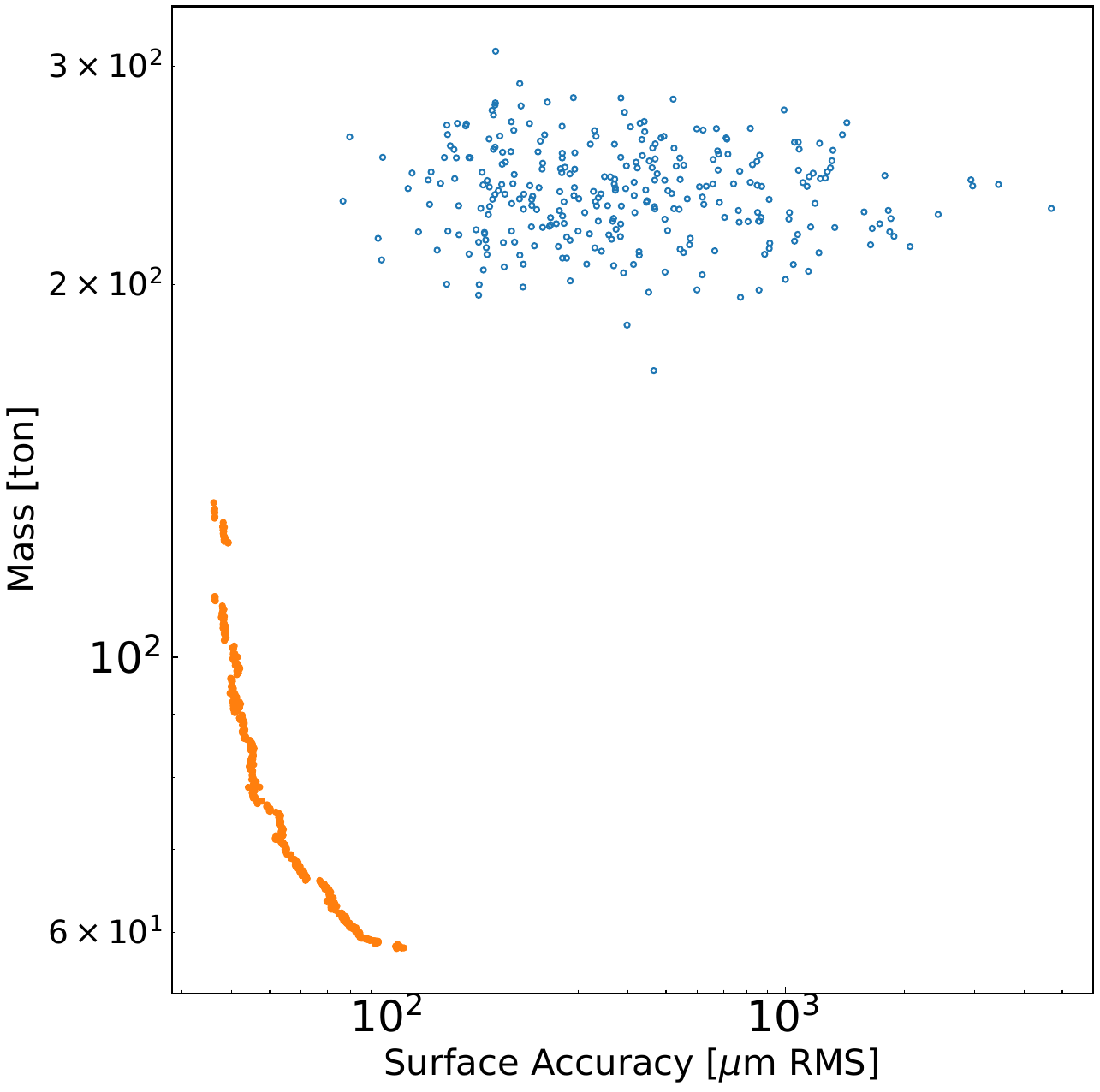}
    \end{center}
    \caption{Randomly generated structures and optimized structures in the surface accuracy and mass of the structure plane in the non-axisymmetric case. Blue circles represent randomly generated initial generation and solid orange circles show final generation. {Alt text: Scatter graph comparing surface accuracy and BUS mass between unoptimized and optimized structures.}}
    \label{fig:asym_pareto_acc_mass}
\end{figure}

\begin{figure*}[tbp]
    \begin{center}
        \includegraphics[width=18cm]{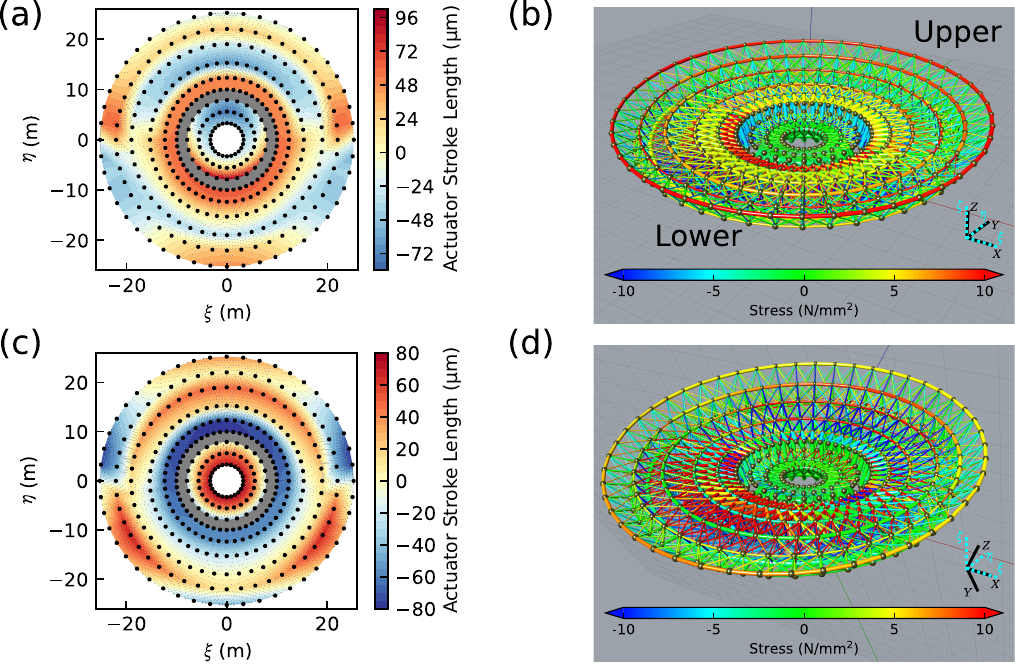}
    \end{center}
    \caption{Maps of actuator stroke length $D_\theta$ and the deformed structures of the optimized structure with the most accurate surface in the non-axisymmetric case. (a) Distribution of actuator stroke lengths at an EL of \ang{85}. (b) Deformed BUS at an EL of \ang{85}. (c) Distribution of actuator stroke lengths at an EL of \ang{30}. (d) Deformed BUS at an EL of \ang{30}. The dotted lightblue axes of (b) and (d) are the local coordinate fixed to the reflector; The solid black axes are the global coordinate. The thickness of the straight truss elements in panels (b) and (d) is exaggerated, and the deformation of the structure is exaggerated by 1000 times. The color of truss elements in panels (b) and (d) corresponds to the stress value: blue indicates compressive stress, and red indicates tensile stress. The colors of truss elements with compressive or tensile stresses over \SI{10}{\newton\milli\meter^{-2}} are saturated. {Alt text: Maps and CAD images of the BUS that exhibits the highest surface accuracy in the non-axisymmetric case, with subfigures labeled a to d, illustrating the structural traits of the BUS.}}
    \label{fig:results_asym}
\end{figure*}

In the optimization of the non-axisymmetric case, results similar to the fiducial case are obtained. Figure~\ref{fig:asym_stats} suggests the convergence of the objective functions to values comparable to those of the fiducial case. Meanwhile, much better structures than those in the fiducial case are not achieved although $\sim10^{70}$ times larger total combinations of truss components than the case in section~\ref{subsec:axially_symmtric}. Given the nature of the GA, it cannot be ruled out that a significantly better solution might be discovered by chance after 100\,000th generation. However, to maintain consistency across the four cases, the search for solutions is terminated at 100\,000th generation.

The surface error and the BUS mass are simultaneously minimized through the simultaneous minimization of the maximum value of the actuator stroke length and BUS mass described in section~\ref{par:objective_functions_and_constraints} using MOGA, as shown in figure~\ref{fig:asym_pareto_acc_mass}. Maximum actuator stroke length, surface accuracy, and mass range from \SI{64.66}{\micro\meter} to \SI{206.8}{\micro\meter}, and from \SI{36.12}{\micro\meter} RMS to \SI{104.6}{\micro\meter} RMS, from 58.24 tons to 133.3 tons, respectively. Surface accuracy and mass are closely correlated, forming a trade-off relationship. Additionally, all optimized structures satisfy the constraints of aperture efficiency and buckling of the truss elements. 

The BUS that exhibits the highest surface accuracy shows structural traits similar to the BUS that exhibits the highest surface accuracy in the fiducial case. At the EL of \ang{85}, the structure is applied with a large stress on the circumferential and radial elements at the ELs of \ang{85} and \ang{30}, respectively, as shown in figure~\ref{fig:results_asym} (b) and (d). The distribution of the straight elements are also similar to the fiducial case. In contrast, the non-axisymmetric design is reflected in the distribution of deformation on the reflector surface, depicted in the panels (a) and (c) of the figure~\ref{fig:results_asym}. The shape of stroke length distributions is arc-shaped. This figure illustrates significant errors near the central hub due to the rigidity around the hub similar to the fiducial case.

\subsection{Axisymmetric Case Optimized Considering the Rigidity of the Central Hub}\label{subsec:sym_wo_hub}

\begin{figure}[tbp]
    \begin{center}
        \includegraphics[width=8cm]{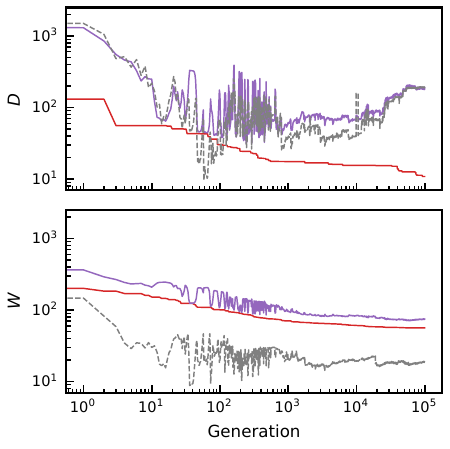}
    \end{center}
    \caption{Statistics in the optimization process of the non-axisymmetric BUS without evaluation of nodes on the central hub. The upper panel shows the objective function of the maximum actuator stroke, and the lower panel shows the objective function of the mass of the BUS. The red line represents the minimum value of the solution, the purple line represents the average, and the gray dashed line depicts the standard deviation. {Alt text: Two line graphs.}}
    \label{fig:sym_wo_hub_stats}
\end{figure}

\begin{figure}[tbp]
    \begin{center}
        \includegraphics[width=8cm]{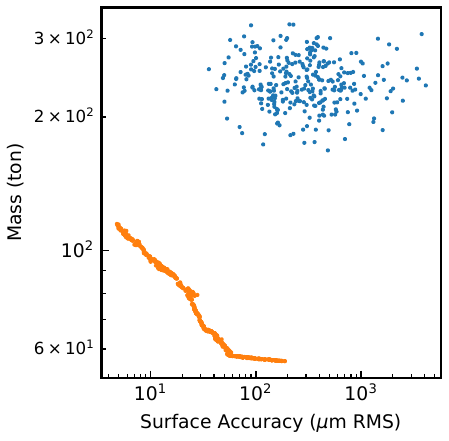}
    \end{center}
    \caption{Randomly generated structures and optimized structures in the surface accuracy and mass of the structure plane in the optimization of axisymmetric case considering the hub rigidity. Blue circles represent randomly generated initial generation and solid orange circles show final generation. {Alt text: Scatter graph comparing surface accuracy and BUS mass between unoptimized and optimized structures.} {Alt text: Scatter graph comparing surface accuracy and BUS mass between unoptimized and optimized structures.}}
    \label{fig:wo_hub_pareto_acc_mass}
\end{figure}

\begin{figure*}[tbp]
    \begin{center}
        \includegraphics[width=18cm]{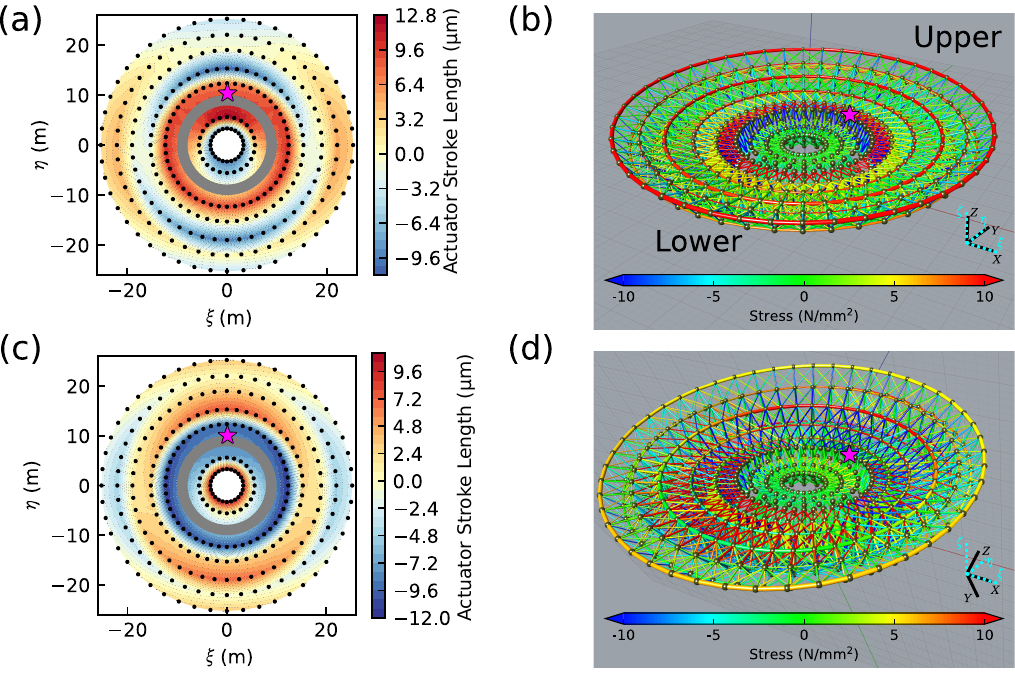}
    \end{center}
    \caption{Maps of actuator stroke length $D_\theta$ and the deformed structures of the optimized structure with the most accurate surface in the axisymmetric case considering the hub rigidity. The star in each panel indicates the node on the central hub requiring the maximum actuator stroke length to correct, which is $\sim +\SI{300}{\micro\meter}$ in this case. (a) Distribution of actuator stroke lengths at an EL of \ang{85}. (b) Deformed BUS at an EL of \ang{85}. (c) Distribution of actuator stroke lengths at an EL of \ang{30}. (d) Deformed BUS at an EL of \ang{30}. The dotted lightblue axes of (b) and (d) are the local coordinate fixed to the reflector; The solid black axes are the global coordinate. The thickness of the straight truss elements in panels (b) and (d) is exaggerated, and the deformation of the structure is exaggerated by 1000 times. The color of truss elements in panels (b) and (d) corresponds to the stress value: blue indicates compressive stress, and red indicates tensile stress. The colors of truss elements with compressive or tensile stresses over \SI{10}{\newton\milli\meter^{-2}} are saturated. {Alt text: Maps and CAD images of the BUS that exhibits the highest surface accuracy in the axisymmetric case considering the hub rigidity, with subfigures labeled a to d, illustrating the structural traits of the BUS.}}
    \label{fig:results_wo_hub}
\end{figure*}

The axisymmetric case considering the hub rigidity are optimized as well. Figure~\ref{fig:sym_wo_hub_stats} depicts the convergence of the objective functions with the diversity of individuals sustained.

The surface accuracy and the BUS mass are simultaneously optimized through the simultaneous minimization of the maximum value of the actuator stroke length and BUS mass described in section~\ref{par:objective_functions_and_constraints} with MOGA as shown in figure~\ref{fig:wo_hub_pareto_acc_mass}. The physical characteristics distribution in the objective function space is analyzed, particularly with respect to the maximum actuator stroke length, surface accuracy, and mass. These range from \SI{10.87}{\micro\meter} to \SI{684.6}{\micro\meter} for maximum actuator stroke length, and from \SI{4.799}{\micro\meter} RMS to \SI{67.91}{\micro\meter} RMS for surface accuracy, from 56.34 tons to 114.8 tons for mass with a distinct trade-off relationship between surface accuracy and mass. We note that nodes on the central hub are ignored in the evaluations of maximum actuator stroke length and surface accuracy as mentioned in section~\ref{subsubsec:methods_wo_central_hub}. The maximum value of the required stroke length of the actuator on the hub is $\sim\SI{300}{\micro\meter}$ at the node indicated as stars in figure~\ref{fig:results_wo_hub} for the structure with the most accurate surface. The large errors around the hub are predictable and can be corrected with a limited number of actuators. These actuators push or pull surface panels with greater stroke lengths than others, which causes them to wear out quickly. However, they are easy to access, and their maintainability is high. In addition, all the optimized structures satisfied both the constraints of aperture efficiency and buckling of the truss elements as in the cases before.

The analysis focuses on the structure with the highest accuracy. Figure~\ref{fig:results_wo_hub} (b) and (d) describe similar structural traits to figure~\ref{fig:results_sym}, but this structure deforms greater than the BUS that exhibits the highest surface accuracy in the fiducial case. Nevertheless, the objective function on the maximum actuator stroke is less than 1/5 of that in the fiducial case. This would be because the algorithm has found the structure which deforms similarly at ELs of \ang{85}, \ang{30}, and \ang{50}, resulting in small $D$. The distribution of deformation on the reflector surface is shown in figure~\ref{fig:results_wo_hub} (a) and (c). The deformation on the whole surface is smaller than that of the fiducial case. Although large errors are found near the central hub, this can be corrected with a large ($\sim \SI{300}{\micro\meter}$) stroke of the actuators around the hub.

\subsection{Non-Axisymmetric Case Optimized Considering the Rigidity of the Central Hub}\label{subsec:asym_wo_hub}

\begin{figure}[tbp]
    \begin{center}
        \includegraphics[width=8cm]{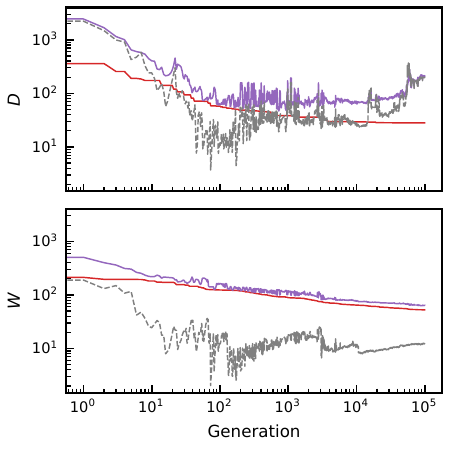}
    \end{center}
    \caption{Statistics in the optimization process of the axisymmetric BUS without evaluation of nodes on the central hub. The upper panel shows the objective function of the maximum actuator stroke, and the lower panel shows the objective function of the mass of the BUS. The red line represents the minimum value of the solution, the purple line represents the average, and the gray dashed line depicts the standard deviation. {Alt text: Two line graphs.}}
    \label{fig:asym_wo_hub_stats}
\end{figure}

The structures in the non-axisymmetric case considering the hub rigidity are also optimized. Figure~\ref{fig:asym_wo_hub_stats} illustrates the convergence of the objective functions while maintaining the diversity of individuals throughout the optimization process. However, similar to section~\ref{subsec:non-axially_asymmetric}, despite the large number of combinations of design variables, the optimization of this case does not yield a significantly better solution compared to its axisymmetric counterpart.

\begin{figure}[tbp]
    \begin{center}
        \includegraphics[width=8cm]{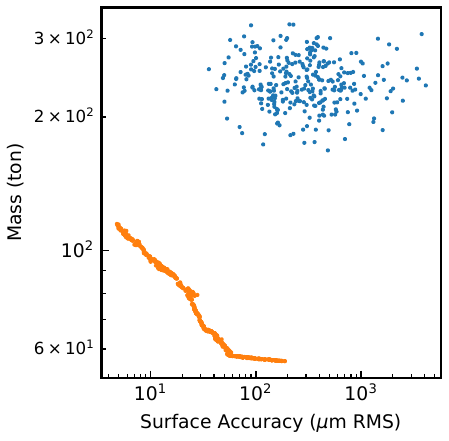}
    \end{center}
    \caption{Randomly generated structures and optimized structures in the surface accuracy and mass of the structure plane in the non-axisymmetric case considering the rigidity of the central hub. Blue circles represent randomly generated initial generation and solid orange circles show final generation. {Alt text: Scatter graph comparing surface accuracy and BUS mass between unoptimized and optimized structures.}}
    \label{fig:wo_hub_asym_pareto_acc_mass}
\end{figure}

\begin{figure*}[tbp]
    \begin{center}
        \includegraphics[width=18cm]{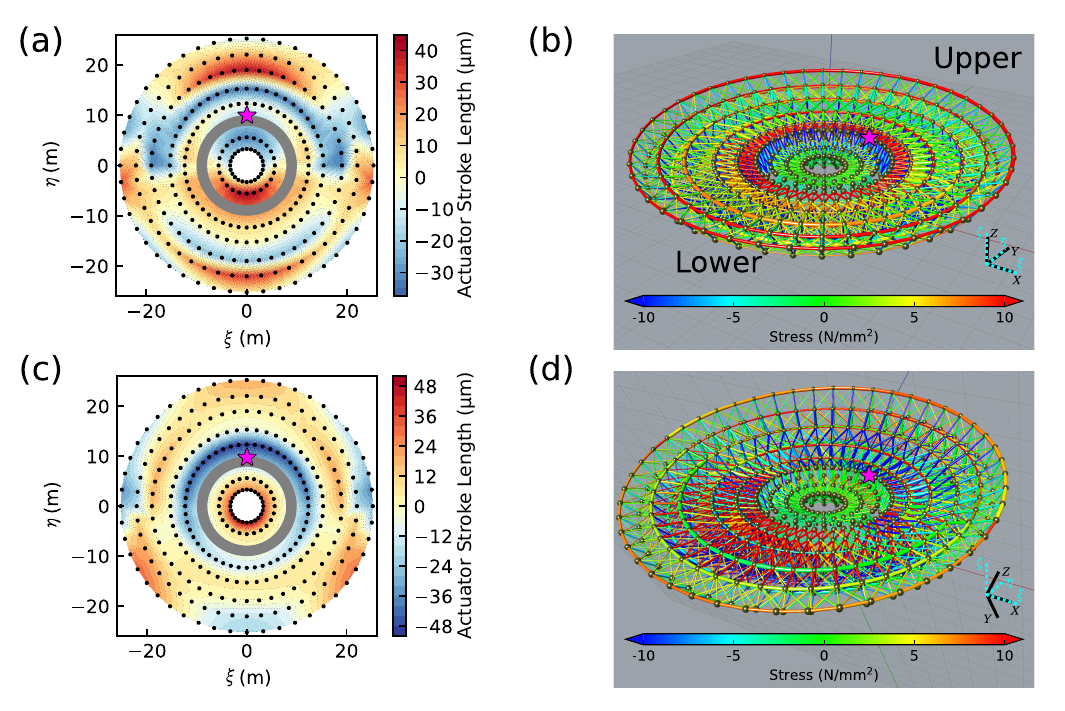}
    \end{center}
    \caption{Maps of actuator stroke length $D_\theta$ and the deformed structures of the optimized structure with the most accurate surface in the non-axisymmetric case considering the hub rigidity. The star in each panel indicates the node on the central hub requiring the maximum actuator stroke length to correct, which is $\sim -\SI{200}{\micro\meter}$ in this case. (a) Distribution of actuator stroke lengths at an EL of \ang{85}. (b) Deformed BUS at an EL of \ang{85}. (c) Distribution of actuator stroke lengths at an EL of \ang{30}. (d) Deformed BUS at an EL of \ang{30}. The dotted lightblue axes of (b) and (d) are the local coordinate fixed to the reflector; The solid black axes are the global coordinate. 
    The width of the straight truss elements in panels (b) and (d) is exaggerated, and the deformation of the structure is exaggerated by 1000 times. The color of truss elements in panels (b) and (d) corresponds to the stress value: blue indicates compressive stress, and red indicates tensile stress. The colors of truss elements with compressive or tensile stresses over \SI{10}{\newton\milli\meter^{-2}} are saturated. {Alt text: Maps and CAD images of the BUS that exhibits the highest surface accuracy in the non-axisymmetric case considering the hub rigidity, with subfigures labeled a to d, illustrating the structural traits of the BUS.}}
    \label{fig:results_wo_hub_asym}
\end{figure*}

The surface accuracy and the BUS mass were simultaneously optimized through the simultaneous minimization of the maximum value of the actuator stroke length and BUS mass described in section~\ref{par:objective_functions_and_constraints} using MOGA, as shown in figure~\ref{fig:wo_hub_asym_pareto_acc_mass}. The physical characteristics distribution within the objective function space was analyzed, focusing on the maximum actuator stroke length, surface accuracy, and mass. These ranged from \SI{27.99}{\micro\meter} to \SI{826.6}{\micro\meter} for maximum actuator stroke length, and from \SI{15.75}{\micro\meter} RMS to \SI{249.3}{\micro\meter} RMS for surface accuracy, with mass varying from 52.80 tons to 92.30 tons. A clear trade-off relationship between surface accuracy and mass is observed. It is important to note that the nodes on the central hub were excluded from the evaluations of the maximum actuator stroke length and surface accuracy, as well as the case of section~\ref{subsec:sym_wo_hub}. For the structure with the highest surface accuracy, the maximum required actuator stroke length on the hub is approximately \SI{200}{\micro\meter} at the node indicated as stars in the figure~\ref{fig:results_wo_hub_asym}. Additionally, all optimized structures satisfies the constraints for aperture efficiency and buckling of the truss elements, as observed in the cases before.

The analysis then focused on the structure with the highest surface accuracy. figure\ref{fig:results_wo_hub_asym} (b) and (d) depict similar structural traits to those in optimized cases before, but this structure exhibited greater deformation compared with the structure with the most accurate surface in section~\ref{subsubsec:methods_non-axisymmetric_structure}. Nevertheless, the objective function for the maximum actuator stroke was less than that in the model of section~\ref{subsubsec:methods_non-axisymmetric_structure}. In summary, as discussed in section~\ref{subsec:sym_wo_hub}, the rigidity of the hub potentially hampers the appropriate homologous deformation mode. The distribution of deformation on the reflector surface is shown in figure~\ref{fig:results_wo_hub_asym} (a) and (c), where the overall surface deformation is smaller than that in the optimized structure in section~\ref{subsubsec:methods_non-axisymmetric_structure}. Although large errors were observed near the central hub, they could be corrected with large actuator strokes (approximately \SI{200}{\micro\meter}) around the hub.

\subsection{Comparison Between Four Cases}\label{subsec:comparison} 

\begin{figure}[!tbp]
    \begin{center}
        \includegraphics[width=8cm]{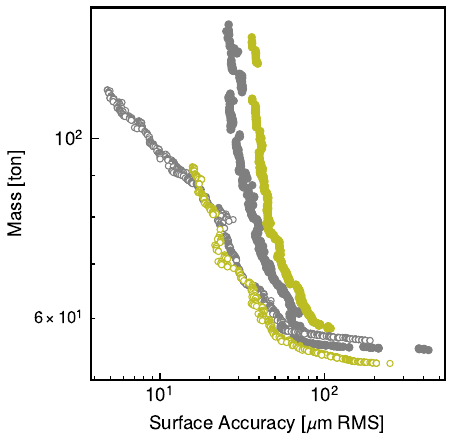}
    \end{center}
    \caption{Comparison of four optimization cases. The gray filled circles represent the axisymmetric case considering the nodes on the central hub in the optimization, the yellow filled circles represent the non-axisymmetric case considering the nodes on the central hub. Open circle represent the optimized structures without considering the nodes on the central hub. The gray open circles represent the axisymmetric case ignoring the nodes on the central hub, and the yellow open circles represent the non-axisymmetric case ignoring the nodes on the central hub. {Alt text: Scatter graph.}}
    \label{fig:comparison}
\end{figure}

In this subsection, we compare the four cases in terms of objective function and structural properties. Figure~\ref{fig:comparison} compares the structure of the four cases in the plane spanned by the surface accuracy and the BUS mass. First, when comparing the axisymmetric and non-axisymmetric models, it is noteworthy that the non-axisymmetric models do not much outperform the axisymmetric model, despite the latter potentially having a larger number of viable solutions. This result suggests that the axisymmetric configuration may be more effective for optimizing the structure under the given constraints. One possible reason for the superiority of the axisymmetric model is the large solution space associated with non-axisymmetric designs. The vastness of this solution space could have led to challenges in finding optimal configurations, making the non-axisymmetric model less effective in achieving high-performance solutions than its axisymmetric counterpart.

Furthermore, it is observed that considering the rigidity of the hub significantly improves the surface accuracy across the symmetry around the $\zeta$-axis. This enhancement in surface accuracy underscores the importance of incorporating hub rigidity in the optimization process, as it plays a crucial role in maintaining the structural integrity and optical performance of the telescope. Notably, a structure in the axisymmetric case considering hub rigidity achieves the highest surface accuracy among all cases. This exceptional performance indicates that such a configuration is particularly well-suited for achieving precise reflector surfaces. By introducing this optimal structure as an initial candidate in the genetic algorithm for the non-axisymmetric model, there is potential to discover even better solutions within the expansive solution space that characterizes non-axisymmetric designs.

The findings suggest that, given adequate iterations, the solution space for non-axisymmetric structures indeed offers possibilities for discovering even more effective structural solutions. This supports the potential use of advanced optimization techniques, such as GA, to explore beyond conventional design boundaries, thereby enabling the design of large-aperture submillimeter telescopes with enhanced performance and reduced material costs.

\section{Discussion}\label{sec:discussion}

\subsection{Natural Frequency of the Optimum Structures}\label{subsec:vibrationalprop}

The natural frequency of a telescope structure plays a critical role in determining how effectively the telescope can respond to external disturbances, as it sets an upper limit on the frequency bandwidth of the control system. A low natural frequency leads to reduced controllability, making it difficult for the telescope to reject disturbances with shorter timescales, such as wind gusts or rapid tracking commands, which may result in degraded pointing performance. Although natural frequency is not a direct target in our optimization process due to high computational cost, its significance cannot be overlooked. Here, we evaluate the natural frequency of the first vibration mode.

An additional modal analysis conducted with OpenSeesPy reveals the first natural frequencies and the corresponding modes of five cases: the fiducial case, the non-axisymmetric case, the axisymmetric case considering hub rigidity, the non-axisymmetric case considering hub rigidity, and the unoptimized axisymmetric case. The first natural frequencies are calculated as \SI{8.248}{\hertz}, \SI{4.781}{\hertz}, \SI{5.363}{\hertz}, \SI{6.528}{\hertz}, and \SI{7.760}{\hertz}, respectively. Various first natural vibration modes are observed. For example, the fiducial case exhibits a torsional mode about the azimuth axis. The non-axisymmetric case shows an asynmetic bowing mode similar to that mode in threefold symmetry around the elevation axis. The axisymmetric case with hub rigidity exhibits a bowing mode with twofold symmetry around the elevation axis. The non-axisymmetric case with hub rigidity exhibits an asymmetric tilting mode about the $\xi$-axis. Finally, the unoptimized axisymmetric case shows a bowing mode with twofold symmetry around the elevation axis.

When comparing the BUSes that exhibits the highest surface accuracy in the fiducial case before and after optimization, it is evident that the natural frequency increased for the structure with the highest surface accuracy. Although the unoptimized structure has high first natural frequency the due to the topology of the BUS with densely arranged straight truss elements, the optimization process successfully improved the rigidity of the structure. 

Meanwhile, the optimized structures of the other cases have lower natural frequencies. In particular, the optimized axisymmetric structure considering hub rigidity is worth analyzing because it achieves the most accurate surface. These lower natural frequencies can be attributed to the elasticity of the structures. The objective function $D$ is evaluated based on the primary reflector correction at an EL angle of \ang{50}, rather than deformations themselves at the EL of \ang{85} and \ang{30}. Thus, even if the deformations are significant, homologous deformation at the EL of \ang{50} can effectively minimize $D$, leading to lower natural frequencies. This trade-off to achieve effective homologous deformation results in reduced natural frequencies.

To address this issue, there are several possibilities to consider. Introducing the first natural frequency as the third objective function could prevent the first natural frequency from being low. If the limiting value of the first natural frequency is defined, setting the frequency as a constraint would also be effective. We discuss this method from the viewpoint of the optimization problem in section \ref{subsec:opt_w/_another_objective function}. Another approach is integrating individuals with higher natural frequencies into the initial generation; it may be possible to prevent decline in natural frequency during the optimization process, thereby maintaining or even improving the structural rigidity. We note that this approach is similar to that of \cite{Kurita2010}, who always entered the rigid structure in every generation of optimization using MOGA. The proposed approach could be particularly beneficial for ensuring that optimized structures achieve high surface accuracy retaining sufficient stiffness, which is critical for the overall performance of the telescope. Although the evaluation of natural frequencies was beyond the scope of the present optimization framework, future work should aim to incorporate them more explicitly into the design formulation. In addition, the influence of the first mode on the azimuth or elevation servo performance should be evaluated.

\subsection{Optimization with Another Objective Function/Method}\label{subsec:opt_w/_another_objective function}

\begin{figure}[!tbh]
    \begin{center}
        \includegraphics[width=8cm]{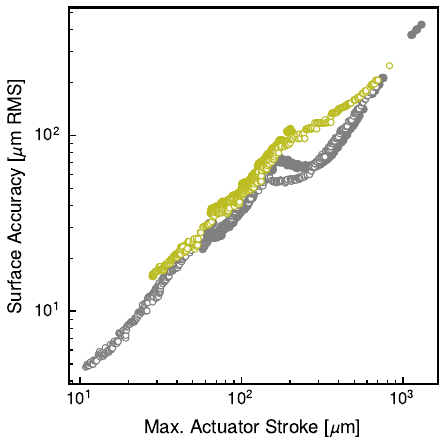}
    \end{center}
    \caption{Comparison between the maximum actuator stroke and surface accuracy. Legend is the same as figure~\ref{fig:comparison}. {Alt text: Scatter graph.}}
    \label{fig:wo_hub_pareto_acc_max}
\end{figure}

Exploring variations in the objective functions can provide insight into the robustness of the optimization process. As an example, we consider the case where one of the objective functions, specifically the maximum actuator stroke length, is replaced by the surface error. When comparing the maximum actuator stroke length with the surface error in the results optimized under the current framework with figure~\ref{fig:wo_hub_pareto_acc_max}, it is evident that these two variables exhibit a very strong correlation. This suggests that similar optimization trends would be expected if the surface error were used as an objective function instead of the maximum actuator stroke length.

Furthermore, considering the discussion in the previous section, it might be tempting to introduce the natural frequency of the structure as a third objective function. However, since the natural frequency is theoretically proportional to the square root of the stiffness-to-mass ratio, incorporating it as an objective function might not be reasonable. Such a choice could lead to conflicting optimization goals, making the optimization process less effective. Conversely, our choice of objective functions would be rational because the choice leads to structures with a high natural frequency. 

Given these considerations, it is unlikely that changing the objective functions in the optimization would significantly impact the results if the variables are strongly correlated. However, selecting correlated variables as objective functions is not advisable, because it may not effectively guide the optimization process. Instead, it is more prudent to choose variables that are either inversely correlated or uncorrelated as objective functions, ensuring that the optimization addresses distinct aspects of the problem and leads to a more balanced and effective design. 

\section{Conclusion}\label{sec:conclusion} 
This study explores the use of Multi-Objective Genetic Algorithms (MOGA) to optimize the BUSes of large-aperture submillimeter telescopes. Our primary objective was to achieve a balance between minimizing surface deformation and reducing the mass of the BUS. Through numerical validation and comparative analysis, we have demonstrated the effectiveness of MOGA in generating optimized structures that challenge traditional design constraints.

Our key findings include:
\begin{itemize}
    \item Our proposed method optimizes the shape and size of \SI{50}{\meter}-class BUS whose best-performing solutions, taken from four different design cases, exhibit maximum actuator stroke lengths ranging from approximately \SI{11}{\micro\meter} to \SI{65}{\micro\meter} (corresponding to \SI{5}{\micro\meter~RMS} to \SI{36}{\micro\meter~RMS} surface error). The formulation of the objective function, in particular, represents a novel contribution, and its effectiveness is demonstrated through numerical validation. 
    \item The optimized non-axisymmetric structures demonstrate an enhanced ability to minimize surface error through the simultaneous minimization of the maximum value of the actuator stroke length and BUS mass while maintaining structural rigidity, although they do not outperform the axisymmetric models as expected because of the vast solution space associated with non-axisymmetric designs.
    \item The axisymmetric model, particularly when considering the rigidity of the central hub, achieves the highest surface accuracy, highlighting the importance of hub rigidity in structural optimization.
    \item The robustness of our method with respect to the choice of objective functions was evaluated by comparing the maximum actuator stroke length and the surface accuracy, which exhibit a strong correlation. This finding suggests that using either variable leads to similar optimization outcomes, thereby validating our selection of actuator stroke as a practical and efficient surrogate for surface accuracy. Furthermore, by avoiding redundant or strongly correlated objectives, the optimization process remains well-conditioned and effective in exploring meaningful trade-offs.
\end{itemize}

These findings suggest that the MOGA is a powerful tool for structural optimization in the design of large-aperture submillimeter telescopes. The approach offers a promising pathway to meet the stringent requirements of next-generation telescopes, such as LST and AtLAST, achieving a surface accuracy of up to $\sim$\SI{5}{\micro\meter} RMS.

Future research should focus on refining the optimization process by incorporating additional factors such as wind-induced and thermal deformations, which are critical for maintaining structural integrity under operational conditions. Additionally, optimizing the support structures for the secondary mirror will be essential to provide the optimum design of the telescope as a whole. These advancements will be crucial in developing more robust and precise designs, ensuring that the next generation of submillimeter telescopes can meet the increasing demands of astronomical observations.

\begin{ack} 
Data analysis was in part carried out on the Multi-wavelength Data Analysis System operated by the Astronomy Data Center (ADC), National Astronomical Observatory of Japan.
\end{ack}

\section*{Funding}
This work was financially supported by JST SPRING Grant Number JPMJSP2125 and JSPS KAKENHI Grant Numbers 22H04939, 23K20035. 

\clearpage

\bibliographystyle{aasjournal}
\bibliography{optlst}

\end{document}